\documentclass[
 reprint,
showpacs,amsmath,amssymb,
 aps, pra,]{revtex4-1}

\usepackage{graphicx}
\usepackage{dcolumn}
\usepackage{bm}
\usepackage{bigstrut}

\begin{document}

\preprint{APS/123-QED}

\newcommand{\api}{$a^3\Pi$} 
\newcommand{\X}{$X^1\Sigma^+$} 

\title{UV frequency metrology on CO (\api); isotope effects and sensitivity to a variation of the proton-to-electron mass ratio}

\author{Adrian J. de Nijs}
\author{Edcel J. Salumbides}
\author{Kjeld S.E. Eikema}
\author{Wim Ubachs}
\author{Hendrick L. Bethlem}
\affiliation{Institute for Lasers, Life and Biophotonics Amsterdam, VU University, De Boelelaan 1081, 1081 HV Amsterdam, The Netherlands}

\date{\today}

\begin{abstract}
UV frequency metrology has been performed on the \api\ - \X\ (0,0) band of various isotopologues of CO using a frequency-quadrupled injection-seeded narrow-band pulsed Titanium:Sapphire laser referenced to a frequency comb laser. The band origin is determined with an accuracy of 5~MHz ($\delta \nu/\nu = 3 \cdot 10^{-9}$), while the energy differences between rotational levels in the \api\ state are determined with an accuracy of 500~kHz. From these measurements, in combination with previously published radiofrequency and microwave data, a new set of molecular constants is obtained that describes the level structure of the \api\ state of $^{12}$C$^{16}$O and $^{13}$C$^{16}$O with improved accuracy. Transitions in the different isotopologues are well reproduced by scaling the molecular constants of $^{12}$C$^{16}$O via the common mass-scaling rules. Only the value of the band origin could not be scaled, indicative of a breakdown of the Born-Oppenheimer approximation. Our analysis confirms the extreme sensitivity of two-photon microwave transitions between nearly-degenerate rotational levels of different $\Omega$-manifolds for probing a possible variation of the proton-to-electron mass ratio, $\mu=m_p/m_e$, on a laboratory time scale.
\end{abstract}


\maketitle

\section{Introduction}

The \api\ state of CO is one of the most extensively studied triplet states of any molecule.
The transitions connecting the \api\ state to the \X\ ground state were first observed by Cameron in 1926~\cite{Cameron}. Later, the \api\ state of the $^{12}$C$^{16}$O isotopologue was studied using radio frequency (rf)~\cite{Freund,Wicke:1972}, microwave (mw)~\cite{Saykally:1987,Carballo,Wada}, infrared~\cite{Havenith,Davies}, optical~\cite{Effantin:1982} and UV spectroscopy~\cite{Field}. The $^{13}$C$^{16}$O isotopologue was studied using rf~\cite{Gammon:1971} and mw~\cite{Saykally:1987} spectroscopy.

Recently, Bethlem and Ubachs~\cite{Bethlem:2009} identified metastable CO as a probe for detecting a temporal variation of the proton-to-electron mass ratio, $\mu=m_p/m_e$, on a laboratory time scale. Two-photon microwave transitions between nearly-degenerate rotational levels in different $\Pi_{\Omega}$ spin-orbit manifolds were shown to be very sensitive to a possible variation of $\mu$. As a measure of the inherent sensitivity of a transition to a drifting $\mu$, the sensitivity coefficient, $K_{\mu}$, is defined via:

\begin{equation}
\frac{\Delta\nu}{\nu} = K_{\mu} \frac{\Delta\mu}{\mu}.
\label{eq:detect_variation}
\end{equation}

\noindent
Transitions between the $J=8,\, \Omega=0$, the $J=6,\, \Omega=1$ and the $J=4, \, \Omega=2$ levels display sensitivities ranging from $K_{\mu}=-300$ to $+200$~\cite{Bethlem:2009}. For an overview on the topic of varying physical constants, we refer the reader to \cite{Uzan} and \cite{UbachsBuning}.

In this paper, we present high-precision UV measurements of the \api\ - \X\ (0,0) band in CO. In total 38 transitions in all six naturally occurring isotopes have been measured with MHz accuracy. All three $\Omega$-manifolds have been probed, with $J$ up to eight. A comprehensive fit of the optical data combined with previously published rf~\cite{Wicke:1972} and mw~\cite{Carballo,Wada} measurements was performed. The molecular constants found for $^{12}$C$^{16}$O are mass scaled and compared with the measured transitions in other isotopologues.

\section{Level structure of CO}

The \api\ state is the first electronically excited state of CO, lying 6~eV above the \X\ ground state. CO in the \api\ state has two unpaired electrons, leading to a nonzero electronic spin, $\vec{S}$, and orbital angular momentum $\vec{L}$. For low rotational levels, the \api\ state is best described in a Hund's case (a) coupling scheme, with the good quantum numbers $\Lambda$ and $\Sigma$, the projection of $\vec{L}$ and $\vec{S}$ on the molecular axis, respectively. The projections of the total angular momentum $\vec{J}$ on the molecular axis and on the space-fixed axis lead to the good quantum numbers $\Omega$ and $M$. The basis functions are $|n\Lambda \rangle|v\rangle|S\Sigma\rangle|J\Omega M\rangle$, representing the electronic orbital, vibrational, electronic spin and rotational components of the wave function, respectively. For higher rotational levels, the spin decouples from the electronic angular momentum and a Hund's case (b) coupling scheme becomes more appropriate. In Hund's case (b), the different $\Omega$-manifolds are mixed.

\begin{figure}
\includegraphics[width=\linewidth]{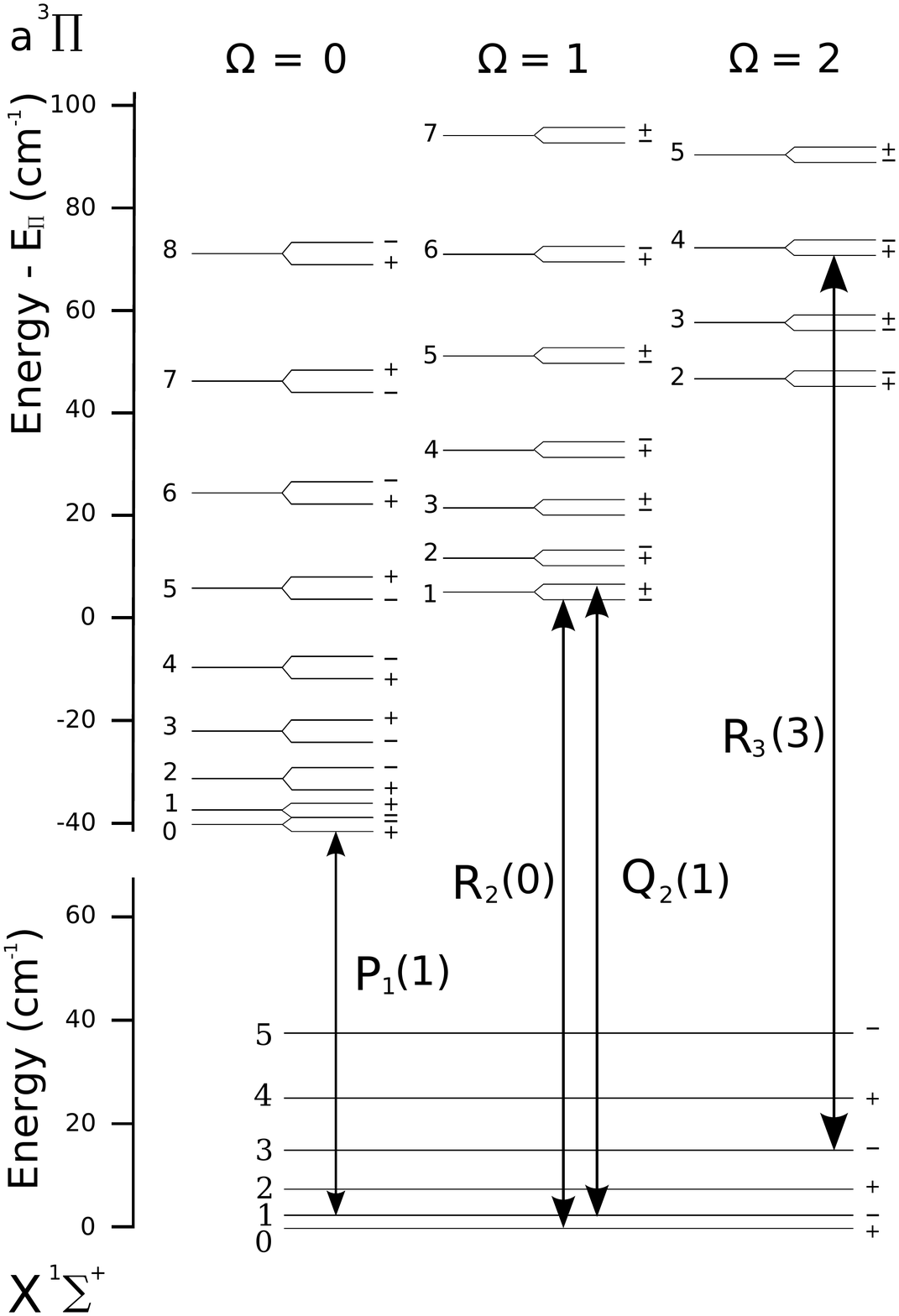}
\caption{Energy level diagram of the \X\ ($v=0$) ground state and the \api\ ($v=0$) state of $^{12}$C$^{16}$O. Rotational quantum numbers and total parity are listed for each level. For the \api\ state, the value of the band origin, $E_\Pi$, is subtracted from the energy scale. The \api\ state has three $\Omega$-manifolds, arising from spin-orbit coupling, and shows lambda-type doubling, as illustrated, not to scale. A number of transitions are indicated by the vertical arrows.}
\label{fig:lvlscheme}
\end{figure}

The energies of the lower rotational levels of the \api\ and \X\ states are shown in Fig. \ref{fig:lvlscheme}, together with selected transitions. The transitions are denoted by $\Delta J_{\Omega+1}(J")$, where transitions with $\Delta J=-1,0$ and $1$ are denoted by $P$, $Q$ and $R$, respectively. As the parity changes in a one-photon transition, the upper lambda-doublet component of rotational levels in the \api\ can only be reached via $Q$-transitions, whereas the lower lambda-doublet components can only be reached via $P$ or $R$ transitions.

The separation of the electronic motion and nuclear motion is not exact, leading to a splitting into lambda-doublet states of opposite parity, as indicated, not to scale, in Fig. \ref{fig:lvlscheme}. The lambda doubling in the $\Omega=0$ state is large and relatively independent of $J$. The lambda doubling in the $\Omega=1$ and $\Omega=2$ manifolds is much smaller. In Fig.~\ref{fig:lvlscheme} the total parity, \emph{i.e.}, the product of the symmetries of the rotational and electronic parts of the wavefunction, is indicated by the - and + signs. The electronic part of the wave function of the upper (lower) lambda-doublet levels has $f$ ($e$) symmetry.

The spin-forbidden \api\ - \X\ system becomes weakly allowed due to spin-orbit mixing of singlet electronic character into \api, most significantly of the $A^1\Pi$ state lying 2~eV above the \api\ state. As the $A^1\Pi$ state consists of a single, $\Omega=1$, manifold, it only couples to the $\Omega=1$ levels in \api. Transitions to the $\Omega=0, J>0$ and $\Omega=2$ manifolds become weakly allowed by mixing of the different $\Omega$-manifolds. The $\Omega=0,J=0$ level is not mixed with the other $\Omega$-manifolds, hence, the transition to this level, \emph{i.e.} the $P_1(1)$, does not obtain transition strength via coupling to the $A^1\Pi$ state.

The \api\ ($v=0$) state can only decay to the ground state, hence, the radiative lifetimes of the different rotational are inversely proportional to the transition strengths. The lifetimes are thus strongly dependent on $J$ and $\Omega$. For example the $J=2,\, \Omega=2$ level has a lifetime of 140~ms, whereas the $J=1, \, \Omega=1$ level has a lifetime of 2.6~ms~\cite{Gilijamse}.

Isotopes with an odd number of nucleons have a non-zero nuclear spin that leads to hyperfine structure. The two relevant odd-nucleon-number nuclei are $^{13}$C and $^{17}$O, with nuclear spin $I=1/2$ and $I=5/2$, respectively. Due to its zero electronic angular and spin momentum the hyperfine splitting in the ground-state of CO is small ($\approx$50~kHz)~\cite{Klapper:2000}. The hyperfine splittings in the \api\ state vary between 30 and 500~MHz for the measured transitions.

\section{Experimental setup}

\begin{figure}
\begin{center}
\includegraphics[width=\linewidth]{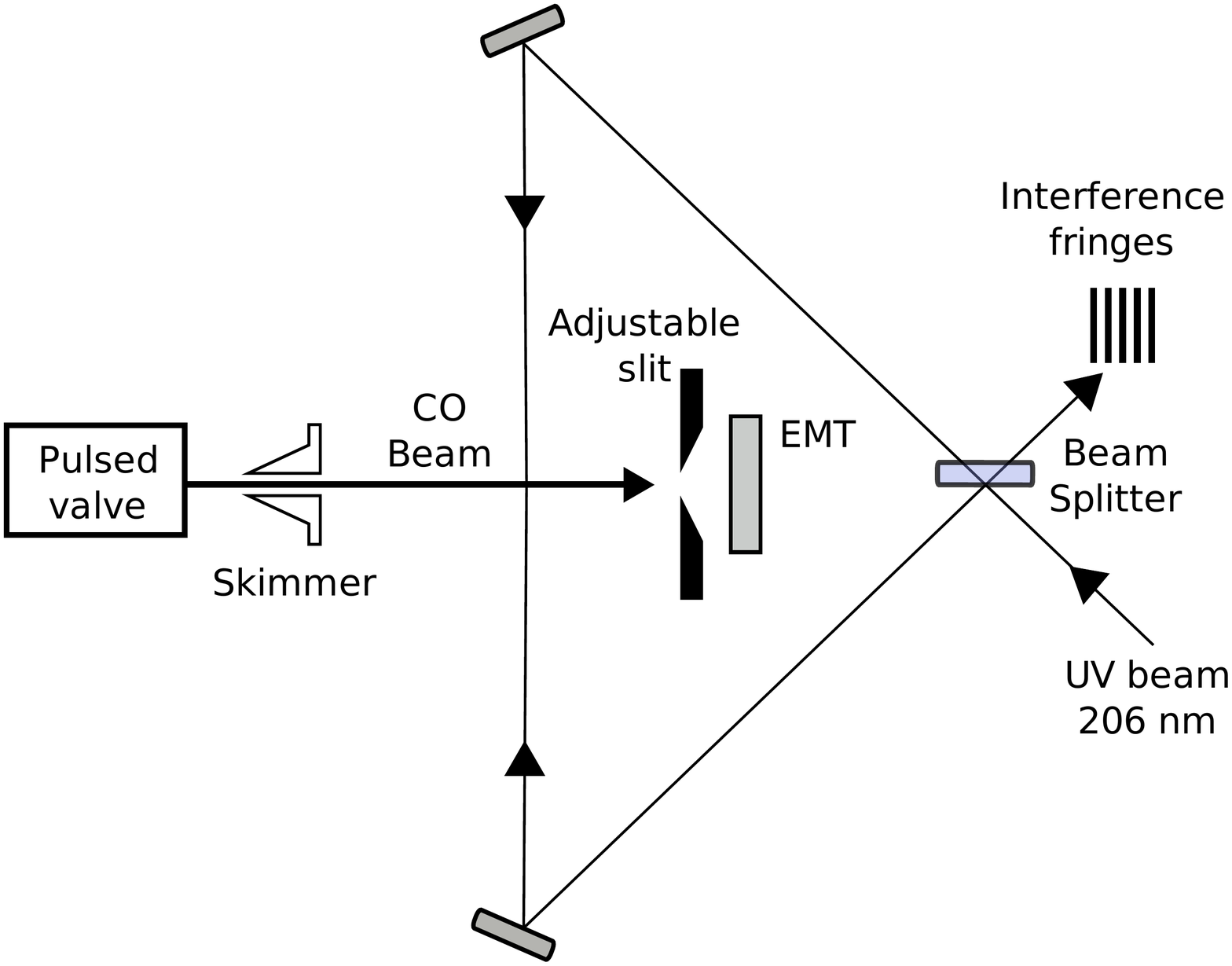}
\end{center}
\caption{A schematic drawing of the experimental setup. A pulsed beam of CO is produced by expanding CO gas into vacuum, using a solenoid valve. After passing through a 1~mm skimmer, the molecular beam is crossed at right angles with laser radiation at 206~nm. After being excited to the \api\ state, the molecules fly 60~cm downstream before hitting an electron multiplier tube, where they are detected. The laser interaction region is built in a Sagnac interferometer to correct for Doppler shifts.}
\label{fig:sagnac}
\end{figure}

The molecular beam setup used for frequency metrology on CO is schematically depicted in Fig.~\ref{fig:sagnac}. A pulsed beam of CO is produced by expanding CO gas into vacuum, using a solenoid valve (General Valve series 9). A backing pressure of 2~bar was used for recording transitions at low $J$, while for recording transitions at higher $J$, the backing pressure was reduced to 0.5~bar. For recording transitions in $^{13}$C$^{16}$O, isotopically enriched CO (Linde Gas) was used. The enriched sample also contained slightly enhanced fractions of $^{13}$C$^{18}$O and $^{13}$C$^{17}$O, sufficient for obtaining signals. Spectra of $^{12}$C$^{17}$O and $^{12}$C$^{18}$O were measured using a natural CO sample. 

After passing through a 1~mm skimmer, the molecular beam is crossed at right angles with laser radiation tunable near 206~nm. In the interaction region, a magnetic field of up to 200~Gauss can be applied by two coils in Helmholtz configuration. After being excited to the \api\ state, the molecules fly 60~cm downstream before being detected by an Electron Multiplier Tube (EMT). The resulting signal is recorded using a digital oscilloscope and the integrated signal is stored. The absolute detection efficiency of this method for metastable CO (6~eV internal energy) is estimated to be on the order of 10$^{-3}$~\cite{Jongma:JCP}. Note that the time of flight (750~$\mu$s) is short compared to the lifetime of the metastable state ($>$2.6~ms). An adjustable slit is mounted in front of the EMT to limit the divergence of the beam that reaches the detector, thereby limiting the Doppler width of the recorded transitions. 

\begin{figure}
\begin{center}
\includegraphics[width=\linewidth]{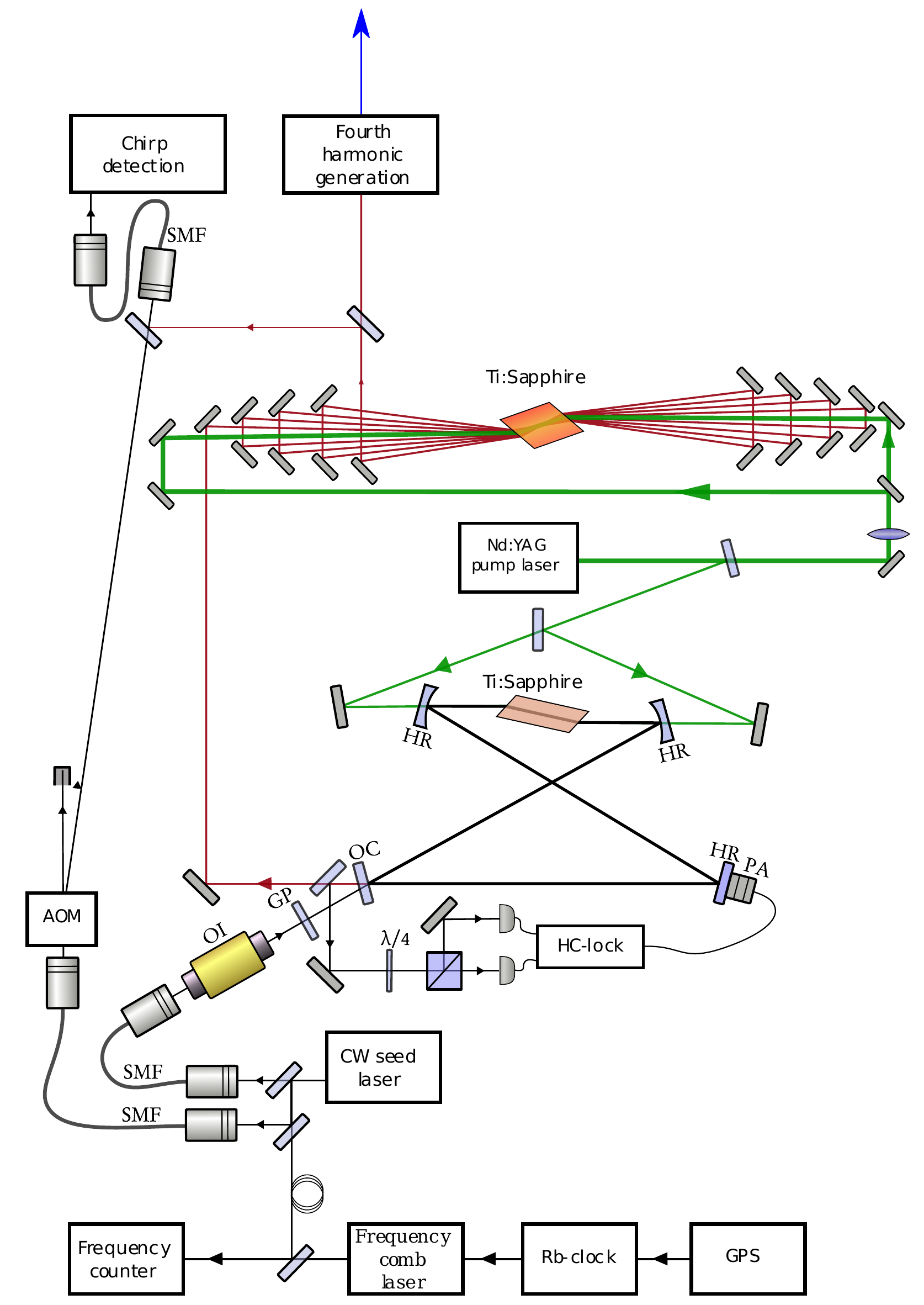}
\end{center}
\caption{(Color online) A schematic drawing of the laser system. A ring laser is used to injection seed an oscillator cavity, which is pumped by a Nd:YAG laser at 10 Hz. The produced IR pulses are amplified in a multi-pass amplifier and subsequently quadrupled in two consecutive BBO crystals. The absolute frequency of the CW seed laser is determined using a fiber comb laser. OI: optical isolator; GP: glass plate; OC: output coupler; HR: high reflective mirror; PA: Piezo Actuator; HC-lock: H\"{a}nsch-Couillaud lock; SMF: single mode fiber; AOM: acousto-optical modulator.}
\label{fig:lasersetup}
\end{figure}

The spectroscopic measurements are performed with a narrow-band frequency-quadrupled titanium:sapphire (Ti:Sa) pulsed laser described in detail by Hannemann \emph{et al.}~\cite{Hannemann:PRA1, Hannemann:PRA2}. A schematic drawing of the laser setup is shown in Fig.~\ref{fig:lasersetup}. A continuous wave (CW) Ti:Sa ring laser (Coherent 899) produces around 700 mW of laser power tunable near 824~nm. Its output is split into three parts, with all parts having approximately the same power. One part of the light is sent as a seed frequency to a pulsed Ti:Sa ring oscillator that is pumped at 10~Hz with 50~mJ of pulsed 532~nm light from an injection seeded Nd:YAG laser (Spectra Physics LAB-170). The oscillator is locked to the CW seed light using a H\"{a}nsch-Couillaud scheme. The pulsed IR light emanating from the oscillator is amplified in a bow-tie multi-pass Ti:Sa amplifier pumped with 300~mJ of pulsed 532~nm light from the same Nd:YAG laser that pumps the oscillator. After nine passes the laser power of the IR-beam is around 70 mJ in a 100 ns pulse. These pulses are then frequency doubled twice in two consecutive BBO crystals, resulting in pulses of 20~$\mu$J at 206~nm.

In order to determine the absolute frequency of the CW Ti:Sa ring laser, the CW light is mixed with the light from 
an erbium-doped fiber frequency-comb laser (Menlo systems MComb at 250 MHz repetition frequency) that is locked to a global positioning system (GPS) disciplined Rb-clock standard. The optical interference beat signal is measured with a photodiode and an Agilent 53132A counter. The obtained beat frequency is then transferred via ethernet to the central computer at which the data is analyzed. Further details on the absolute frequency calibration can be found in Sec.~\ref{subsubsec:comb}.

By making a small portion of the pulsed light interfere with part of the CW light, possible small differences between the frequencies of the CW seed laser and the central frequency of the pulsed output of the bow-tie amplifier are measured and corrected for~\cite{Hannemann:PRA1}. In order to have a good fringe visibility, and to ascertain that the full wavefront of the pulsed output is mapped onto the CW reference beam, both beams are sent through a short single-mode fiber. The beat pattern is detected using a fast photodiode in combination with an oscilloscope and analyzed online.

The UV laser beam is split into two parts and sent through the molecular beam machine from opposite sides to limit the Doppler shift due to a possible imperfect perpendicular alignment of the laser beam. In order to ensure that the two beams are perfectly counterpropagating, the two laser beams are recombined after passing through the machine, forming a Sagnac interferometer. The paths through the molecular beam machine are aligned such that the two beams interfere destructively at the exit port (a dark fringe). The transition frequency is measured twice, using either the laser beam from the left or from the right hand side~\cite{Hannemann:OptLett}.

\section{Experimental results}

\begin{figure}
\begin{center}
\includegraphics[width=1.1\linewidth]{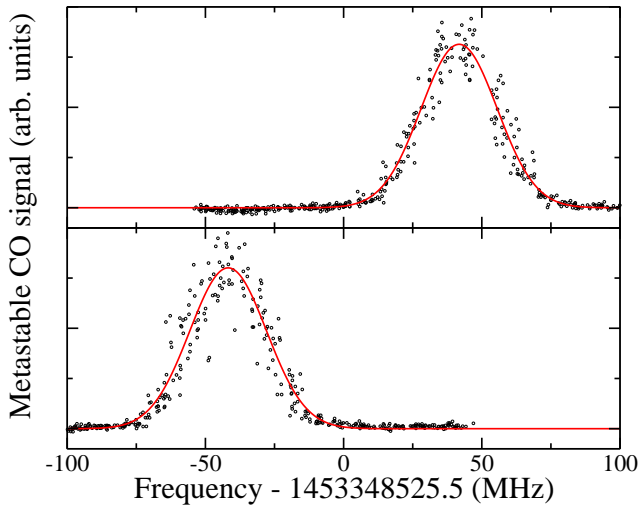}
\end{center}
\caption{(Color online) Recordings of the $R_2(0)$ transition in $^{12}$C$^{16}$O measured using the laser beam propagating from the right hand side, upper panel, and the left hand side, lower panel. The frequency of the seed laser is scanned while the signal from the EMT is recorded. Each point represents a single laser pulse. For each pulse the beat frequency of both the fiber comb and the CW-pulse offset measurement setup are recorded. A single scan takes around 20 minutes.}
\label{fig:typical}
\end{figure}

\begin{table}
\begin{footnotesize}
\begin{ruledtabular}\begin{tabular}{lrrrr}
Isotopologue&Transition&Observed (MHz)&Residuals&Residuals\\
&&&Fitted&Scaled\\
\hline
\bigstrut
$^{12}$C$^{16}$O	&	$	R_1(0)	$	&	1452~065~305.5	&	0.4	&	---	\\ 
	&	$	P_1(1)	$	&	1451~857~145.0	&	0.9	&	---	\\
	&	$	Q_1(1)	$	&	1452~002~039.3	&	1.3	&	---	\\
	&	$	R_1(7)	$	&	1452~112~469.1	&	2.1	&	---	\\
	&	$	Q_1(8)	$	&	1451~232~745.8	&	2.4	&	---	\\
	&	$	R_2(0)	$	&	1453~348~525.5	&	1.1	&	---	\\
	&	$	Q_2(1)	$	&	1453~233~648.6	&	1.2	&	---	\\
	&	$	R_2(5)	$	&	1453~607~338.8	&	-4.8	&	---	\\
	&	$	Q_2(6)	$	&	1452~922~384.0	&	-6.6	&	---	\\
	&	$	P_2(7)	$	&	1452~109~212.5	&	-6.2	&	---	\\
	&	$	R_3(1)	$	&	1454~488~881.8	&	1.8	&	---	\\
	&	$	Q_3(2)	$	&	1454~258~350.7	&	1.4	&	---	\\
	&	$	R_3(3)	$	&	1454~683~187.7	&	2.2	&	---	\\
	&	$	Q_3(4)	$	&	1454~222~240.4	&	2.7	&	---	\\
\hline	
\bigstrut
$^{12}$C$^{17}$O	&	$	R_2(0)	$	&	1453~426~677.0	&	---&	---\\
	&	$		$	&	1453~426~924.0&	---&	---\\
	&	$		$	&	1453~427~108.2&	---&	---\\
\hline	
\bigstrut									
$^{12}$C$^{18}$O	&	$	R_2(0)	$	&	1453~498~161.1&	---&	---\\ 
\hline	
\bigstrut									
$^{13}$C$^{16}$O	&	$	R_1(2)	$	&	1452~325~025.5	&	-0.2	&	0.2	\\ 
	&	$		$	&	1452~325~217.9	&	-0.3	&	0.7	\\
	&	$	R_1(7)	$	&	1452~263~496.1	&	-1.6	&	-13.1	\\
	&	$		$	&	1452~263~906.8	&	2.3	&	-8.0	\\
 	&	$	Q_1(8)	$	&	1451~425~012.3	&	-0.3	&	-9.9	\\
	&	$		$	&	1451~425~321.6	&	0.5	&	-10.1	\\
	&	$	R_2(0)	$	&	1453~491~411.3	&	0.8	&	3.4	\\
	&	$		$	&	1453~491~449.2	&	1.8	&	4.4	\\
	&	$	Q_2(1)	$	&	1453~381~555.9	&	0.4	&	2.5	\\
	&	$		$	&	1453~381~613.6	&	-0.2	&	2.4	\\
	&	$	R_2(1)	$	&	1453~570~841.9	&	-0.5	&	3.8	\\
	&	$		$	&	1453~570~876.4	&	-0.6	&	3.4	\\
	&	$	R_2(5)	$	&	1453~740~818.2	&	0.7	&	5.2	\\
	&	$		$	&	1453~740~854.3	&	-0.9	&	2.8	\\
	&	$	Q_2(6)	$	&	1453~085~785.3	&	-0.6	&	0.1	\\
	&	$		$	&	1453~085~898.7	&	-0.6	&	0.9	\\
	&	$	P_2(7)	$	&	1452~308~565.9	&	-0.3	&	4.2	\\
	&	$		$	&	1452~308~602.9	&	-0.9	&	2.8	\\
	&	$	R_3(1)	$	&	1454~635~451.9	&	-1.2	&	-3.1	\\
	&	$		$	&	1454~636~019.2	&	-1.0	&	-2.7	\\
	&	$	R_3(3)	$	&	1454~818~557.3	&	0.9	&	2.8	\\
	&	$		$	&	1454~818~676.7	&	0.8	&	2.6	\\
	&	$	Q_3(4)	$	&	1454~377~868.2	&	0.8	&	2.6	\\
	&	$		$	&	1454~377~989.2	&	0.1	&	2.0	\\
\hline
\bigstrut			
$^{13}$C$^{17}$O	&	$	R_2(0)	$	&	1453~571~485.7	&	---&	---\\ 
	&	$		$	&	1453~571~732.5&	---&	---\\ 
	&	$		$	&	1453~571~918.5&	---&	---\\ 
\hline	
\bigstrut								
$^{13}$C$^{18}$O	&	$	Q_1(3)	$	&	1452~210~864.9	&	---	&	-12.3	\\ 
	&	$		$	&	1452~211~002.6	&	---	&	-12.4	\\
	&	$	R_2(0)	$	&	1453~643~585.2	&	---	&	-1.7	\\
	&	$		$	&	1453~643~626.6	&	---	&	-1.4	\\
	&	$	Q_2(1)	$	&	1453~539~184.1	&	---	&	-4.4	\\
	&	$		$	&	1453~539~244.9	&	---	&	-4.6	\\
	&	$	R_2(5)	$	&	1453~882~578.9	&	---	&	15.9	\\
	&	$		$	&	1453~882~612.1	&	---	&	13.8	\\
	&	$	Q_2(6)	$	&	1453~259~995.8	&	---	&	7.6	\\
	&	$		$	&	1453~260~099.7	&	---	&	4.4	\\
	&	$	R_3(3)	$	&	1454~963~029.8	&	---	&	-0.5	\\
	&	$		$	&	1454~963~165.2	&	---	&	1.7	\\
	&	$	Q_3(4)	$	&	1454~544~280.2	&	---	&	-2.6	\\
	&	$		$	&	1454~544~414.5	&	---	&	-3.4	\\
\end{tabular}
\end{ruledtabular}
\caption{Measured transition frequencies for various isotopologues of CO. The rightmost two columns are the residuals from the fit and the mass-scaling procedure discussed in Sec.~\ref{sec:ana}. \label{tab:fitresults}} 
\end{footnotesize}
\end{table}

In Table~\ref{tab:fitresults} the frequencies are listed for the measured transitions in the \api\ - \X\ (0,0) band. In view of the time-consuming measurement procedure, only a selection of transitions has been investigated. In total, 38 transitions have been recorded in all six stable isotopologues of CO. In $^{12}$C$^{16}$O, $^{13}$C$^{16}$O and $^{13}$C$^{18}$O, we have recorded low $J$ transitions to each $\Omega$-manifold and each parity level, to be able to give a full analysis of the level structure. Also, transitions to the $J=8,\,\Omega=0$, $J=6,\,\Omega=1$ and $J=4,\,\Omega=2$, the near degenerate levels of interest to the search for $\mu$ variation, were measured. Furthermore, the $R_2(0)$ transition, the most intense line under our conditions, has been measured in the $^{12}$C$^{17}$O, $^{12}$C$^{18}$O and $^{13}$C$^{17}$O isotopologues. Note that the observed signal strengths for the different transitions varies over more than three orders in magnitude.

In Fig.~\ref{fig:typical} a typical recording of the $R_2(0)$ transition in $^{12}$C$^{16}$O is shown. The upper and lower graph show the spectra obtained with the laser beam propagating through either path of the Sagnac interferometer. For each pulse of the pulsed laser, the frequency of the CW laser is determined using the fiber comb laser and a possible shift between the pulsed laser and the CW laser is determined using the online CW-pulse offset detection. Subsequently, these data are combined with the metastable CO signal from the EMT. Typically, the recorded scans are not perfectly linear, resulting in an uneven distribution of the data points along the frequency-axis. This has no influence on the peak determination. At the peak of the transition, the observed signal corresponds to typically a few thousand detected metastable CO molecules per laser pulse. Note that the fluctuations in the signal shown in Fig.~\ref{fig:typical} are due to pulse-to-pulse variations of the molecular beam and the UV power and not due to counting (Poisson) statistics. The solid lines in the figure show a Gaussian fit to the spectra. The transition frequency of the $R_2(0)$ transition is determined by taking the average of the measurements taken with the laser beam propagating from either side.

\begin{figure}
\begin{center}
\includegraphics[width=1.1\linewidth]{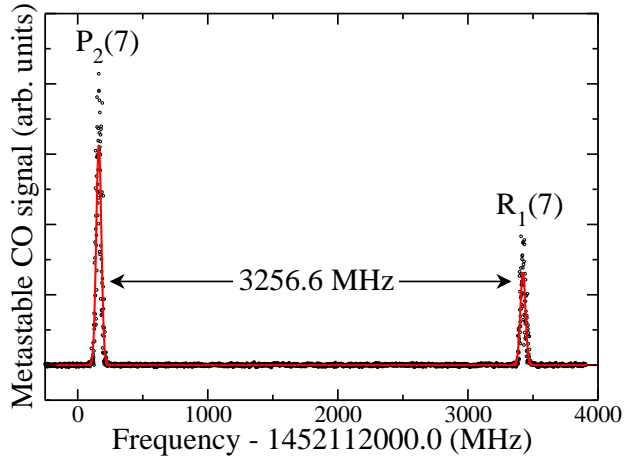}
\end{center}
\caption{(Color online) Recording of the $P_2(7)$ transition and the $R_1(7)$ transition in $^{12}$C$^{16}$O. Both transitions originate from the same ground state level and connect to one of two nearly degenerate levels. Therefore, the combination difference of these two transition frequencies corresponds to the frequency of the transition between the two nearly-degenerate levels.}
\label{fig:longscan}
\end{figure}

Transitions to the six near-degenerate levels in both $^{12}$C$^{16}$O and $^{13}$C$^{16}$O were measured. For $^{13}$C$^{18}$O transitions to four of the six near degenerate levels were obtained. In Fig.~\ref{fig:longscan} a recording of the $P_2(7)$ and $R_1(7)$ transitions in $^{12}$C$^{16}$O is shown. Both transitions originate from the $J=7$ ground state level, thus the combination difference is equal to the frequency of the $J=6,\,\Omega =1,\,+ \rightarrow J=8,\,\Omega=0,\,+$ transition, which is measured to be 3256.6 MHz.

As the transition strengths of transitions in the spin forbidden \api\ - \X\ system originate from mixing of \api\  $\Omega=1$ with $A^{1}\Pi$~\cite{Gilijamse}, the transition strengths of the different transitions are proportional to (the square of) the $\Omega=1$ character of the final rotational level and the H\"onl-London factor. From this, we expect the $P_2(7)$ to be 3.2 times more intense than the $R_1(7)$. Experimentally, we find a ratio of 2.4 to 1. The deviation is explained by the fact that the $J=8,\, \Omega=0$ has a longer lifetime than the $J=6,\, \Omega=1$ (15.7~ms vs. 3.5~ms), and consequently, a smaller fraction of the metastable molecules decays back to the ground state before reaching the EMT. Taking the lifetime into account, we expect a ratio of 2.7 to 1, in reasonable agreement with the experiment.

\begin{figure}
\begin{center}
\includegraphics[width=1.1\linewidth]{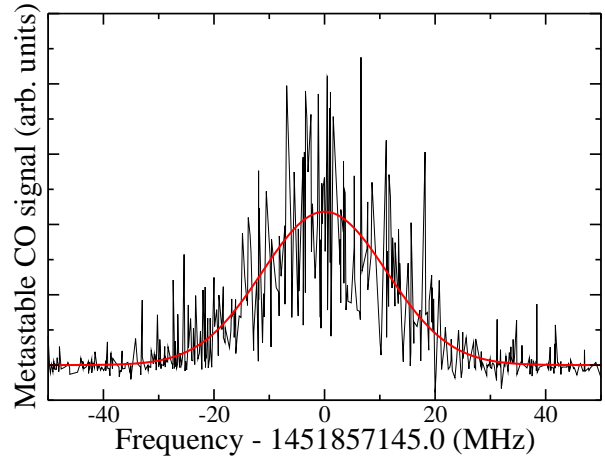}
\end{center}
\caption{(Color online) A recording of the $P_1(1)$ transition in $^{12}$C$^{16}$O. If only the coupling to the $A^1\Pi$ state is considered, this transition has zero transition strength.}
\label{fig:P11}
\end{figure}

The $J=0,\, \Omega=0$ is the only $J=0$ level, and is therefore not mixed with the $\Omega=1$ and $\Omega=2$ manifolds. Consequently, the $P_1(1)$ transition, connecting the \X\ $J=1$ with the \api\ $\Omega=0,J=0$, + parity level does not obtain any transition strength from coupling to the $A^{1}\Pi$ state. Nevertheless, we were able to observe this transition, shown in Fig.~\ref{fig:P11}, albeit with low signal to noise. From our measurements, we estimate that the $P_1(1)$ transition is about 65~times weaker than $Q_1(1)$ and about $10^{4}$ times weaker than $Q_{2}(1)$, resulting in a lifetime of 8(1)~s~\cite{groenenboomprivatecommunications}. The transition strength is ascribed to mixing of the \api\ state with a $^{1}\Sigma^{+}$ state, most likely the \X\ ground state~\cite{Minaev}.

\begin{figure}
\begin{center}
\includegraphics[width=1.1\linewidth]{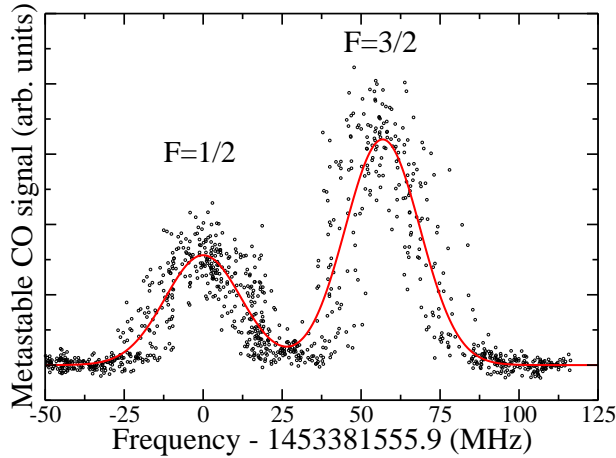}
\end{center}
\caption{(Color online) A recording of the $Q_2(1)$ transition in $^{13}$C$^{16}$O. The labels indicate the value of the total angular momentum $\vec{F}=\vec{J}+ \vec{I}$ in the excited state.}
\label{fig:hyperfine}
\end{figure}

The rotational levels of isotopologues with one or two odd-numbered nuclei show hyperfine splitting. In Fig.~\ref{fig:hyperfine} a recorded spectrum of the $Q_2(1)$ transition in $^{13}$C$^{16}$O is shown. The transitions to the hyperfine sublevels $F=J-I=1/2$ and $F=J+I=3/2$ of the excited state are clearly resolved.

\section{Measurement uncertainties}

In this paragraph the main sources of uncertainty in the transition frequencies are discussed. The uncertainty budget is summarized in Table~\ref{tab:error}.

\subsection{Zeeman effect}

The ground state of CO is a $^1\Sigma$ state, hence, its Zeeman shift is small. The \api\ state on the other hand has both electronic angular momentum and spin, and experiences a considerable Zeeman shift. In the $\Omega=2$ state the effect due to the electronic angular momentum and due to spin will add up, whereas in the $\Omega=0$ and $\Omega=1$ state these effects will partly cancel. Hence, we expect the largest Zeeman shift to occur in the $\Omega=2$ levels, in particular in the $J=2,\Omega=2$ level.

\begin{figure}
\begin{center}
\includegraphics[width=1.1\linewidth]{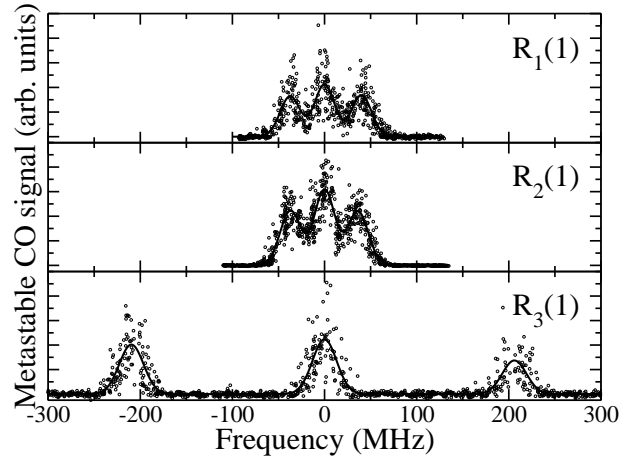}
\end{center}
\caption{The $R_1(1)$, $R_2(1)$ and $R_3(1)$ transitions in $^{12}$C$^{16}$O measured in an applied magnetic field of 170 Gauss, showing Zeeman splitting.}
\label{fig:Zeeman}
\end{figure}

In Fig.~\ref{fig:Zeeman} the recorded spectra are shown for the $R_{1}(1)$, $R_{2}(1)$ and $R_{3}(1)$ transitions in $^{12}$C$^{16}$O in a magnetic field of 170~Gauss. In these measurements, the polarization of the 206~nm light is parallel to the applied magnetic field, hence, only $\Delta M_{J}=0$ transitions are allowed. As all three measured transitions originate from the \X\ $J=1$ level only the $M_J= \pm1,0$ components of the probed \api\ levels are observed. The Zeeman shift of the $J=2,\,\Omega=2,\,M_{J}=1$ state is 1.2~MHz/Gauss, while the shifts of the $J=2,\,\Omega=1,\,M_{J}=1$ and $J=2,\,\Omega=0,\,M_{J}=1$ are about 5 times smaller. It is observed that the transitions to the $M_J=-1$ and $M_J=1$ states are approximately equally strong, implying that the polarization of the 206~nm is nearly perfectly linear, estimated to be better than 99\%, as expected from light that has been quadrupled in non-linear crystals. In the earth magnetic field,  approximately $0.5$~Gauss, the $J=2,\,\Omega=1,\,M_{J}=-2$ and $M_{J}=2$, will be shifted in opposite directions by less than 1~MHz. As the polarization is nearly perfect, the shift of the line center is estimated to be less than 1~kHz.

\subsection{Stark effect}

The Stark shift in the ground state of CO is negligibly small as the dipole moment is only 0.1~Debye (corresponding to $0.17\cdot10^{-2}$cm$^{-1}$/(kV/cm)) and mixing occurs between rotational levels. The Stark shift in the \api\ state on the other hand is larger as the dipole moment is 1.37~Debye and mixing occurs between the lambda-doublet components~\cite{Jongma:CPL}. We estimate that the electric field in the excitation region is below 1~V/cm. This corresponds to a Stark shift of 30~kHz for the $J=2,\,\Omega=2,\,M\Omega=4$ and less for all other levels.

\subsection{AC-Stark effect}

A priori, the AC-Stark shift is difficult to estimate. We have measured the $R_{2}(0)$ transition in $^{12}$C$^{16}$O for different laser powers, reducing the laser power by over an order of magnitude, but found no significant dependence of the transition frequency on laser power. Thus, we estimate the AC-Stark shift to be less than 100~kHz.

\subsection{Uncertainty in absolute frequency determination}
\label{subsubsec:comb}

The absolute frequency of the CW light is calibrated by mixing this light with the output of a frequency-comb laser and counting the resulting beat frequency $f_{\mathrm{beat}}$. The frequency $f_{\mathrm{CW}}$ is then obtained by the relation 

\begin{equation}
f_{\mathrm{CW}} = n \cdot f_{\mathrm{rep}} + f_{0} + f_{\mathrm{beat}}, 
\end{equation}

\noindent with $n$ the mode number of the frequency comb that is nearest to the frequency of the CW light, and $f_{\mathrm{rep}}$ and $f_{0}$ the repetition and the carrier-envelope offset frequencies of the frequency comb, respectively. $f_{\mathrm{rep}}$ is tunable over a small range around 250~MHz and $f_{0}$ is locked at 40~MHz. We infer that the sign of $f_{\mathrm{beat}}$ is positive from the observation that $f_{\mathrm{beat}}$ increases when the CW laser is scanned towards higher frequency. Likewise, the sign of $f_{0}$ is positive from the observation that $f_{\mathrm{beat}}$ increases when $f_{0}$ is decreased.  The beat note is averaged over a period of 100~ms. On this time scale, the accuracy of the Rb-clock standard is $10^{-10}$, equivalent to 150~kHz at the used frequencies. This uncertainty enters separately in each data point taken in a frequency scan. As each scan is approximately 1000 data points, this uncertainty averages out.

The integer mode number $n$ is determined by measuring the transition frequency of the $R_{2}(0)$ transition in $^{12}$C$^{16}$O using the frequency comb at different repetition frequencies and then determining at which transition frequency these three measurements coincide. This method gives an unambiguous transition frequency provided that the change in repetition frequency is much larger than the measurement uncertainty~\cite{Witte:Science}. In our case this condition is well met as the uncertainty between two consecutive measurements is on the order of 500~kHz, while the repetition frequency of the comb is varied by 2~MHz. The absolute value obtained for the $R_{2}(0)$ transition of $^{12}$C$^{16}$O is used to calibrate a wavelength meter (Burleigh WA-1500, 30~MHz precision) on a daily basis. This wavelength meter is then used to determine the mode number for the measurements of the other transitions.

\subsection{Doppler effect}
\label{subsubsec:sagnac}

\begin{figure}
\begin{center}
\includegraphics[width=1.1\linewidth]{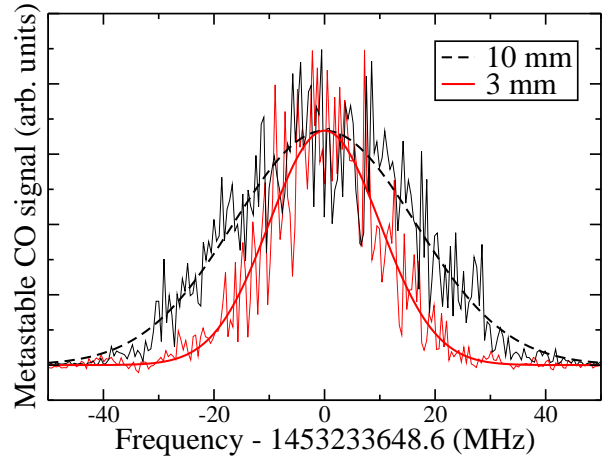}
\end{center}
\caption{(Color online) The $Q_2(1)$ transition measured using slit widths of 10~mm and 3~mm. When the width of the slit is decreased the transition becomes narrower, down to 23~MHz for a slit width of 3~mm. The 23~MHz spectral width is attributed to the line width of the UV-laser.}
\label{fig:dopplerwidth}
\end{figure}

In a molecular beam experiment, the first-order Doppler effect is reduced by aligning the laser-beam perpendicular to the molecular beam. Two residual effects remain: (i) The finite transverse temperature of the molecular beam leads to a broadening of the transition, (ii) A possible imperfect perpendicular alignment of the laser beam leads to a shift of the center frequency.

To limit the Doppler width, we have placed a variable slit in front of the EMT, which limits the divergence of the beam hitting the detector. In Fig.~\ref{fig:dopplerwidth} recordings of the $Q_2(1)$ transition of $^{12}$C$^{16}$O using a slit width of 10~mm and 3~mm are shown. When the slit width is reduced below 3~mm the width of the transition remains the same while the signal decreases further. The minimum full width at half maximum (FWHM) observed is 23~MHz, which is attributed to the line width of the UV-radiation. In our measurements, we have used a slit width of 6~mm, as a compromise between signal intensity and line width.

In order to eliminate the Doppler shift in the measurement, the UV laser beam is aligned in the geometry of a Sagnac interferometer. The angle between the two counterpropagating laser beams may be estimated to be smaller than $\lambda/d$, with $\lambda$ being the wavelength of the light in nm and $d$ being the diameter of the laser beam in mm~\cite{Hannemann:OptLett}. In our case this results in a maximum Doppler shift of 200~kHz. We have verified this by comparing measurements of the $R_2(0)$ transition of $^{12}$C$^{16}$O in a pure beam of CO (velocity of 800~m/s) and a beam of 5\% CO seeded in He (longitudinal velocity of 1500~m/s).
The second-order Doppler shift is sub-kHz. The recoil shift is $\approx$~25~kHz.

\subsection{Uncertainty in peak determination}

The number of metastable molecules detected (at the peak of) the $R_2(0)$ transition of $^{12}$C$^{16}$O, the strongest transition observed, is on the order of $10^{4}$ per laser pulse, while at the $P_1(1)$ transition of $^{12}$C$^{16}$O, the weakest transition observed, it is only a few molecules per laser pulse. The ability to determine the line center from the measurement is limited by pulse-to-pulse variations of the molecular beam and the UV power. Typically, the uncertainty in the peak determination is 300~kHz (corresponding to 1\% of the line-width).

\subsection{Frequency chirp in the pulsed laser system}
\label{subsubsec:chirp}

The frequency of the pulsed IR light differs slightly from the frequency of the CW seed laser mainly due to two effects: (i) Mode pulling in the oscillator and (ii) Frequency chirp and shift in the amplifier. The pump laser induces a change of the refractive index of the Ti:Sa crystal and therefore a variation of the optical path length of the cavity on a time scale much shorter than the response time of the electronic system used for locking the cavity. Therefore, at the instant the pulse is produced the cavity resonances are shifted from the frequency they are locked to. The size of the frequency shift due to mode pulling depends on the pump power. The optical properties of the Ti:Sa crystal in the bow-tie amplifier change depending on the population inversion, which decreases during the multi-step amplification of the IR pulse. The combined effect of both phenomena results in a shift on the order of a few MHz. We measure and compensate for these effects by making a part of the pulsed light interfere with part of the CW light. The CW-pulse offset can be determined within 200~kHz. However, as a result of wavefront distortions of the pulsed beam, it depends very critically on the alignment. We have measured the chirp at different positions in the wavefront by moving a small pinhole through the pulsed beam, and observed deviations of a few MHz. These deviations are amplified in the harmonic generation stages. As it is unclear which part of the pulsed laser beam gives rise to the observed metastable CO signal, we cannot compensate for this effect. Consequently, we find rather large day-to-day deviations in the measurements; measured transition frequencies taken on the same day, with a specific alignment and setting of the Ti:Sa oscillator and bow-tie amplifier, agree within a 500~kHz. Measured transition frequencies taken on different days, on the other hand, may deviate by a few MHz. In order to compensate for this systematic effect, we chose the $R_{2}(0)$ transition in $^{12}$C$^{16}$O as an anchor for each measurement session. With this anchoring, measurements of any given transition (other than the $R_{2}(0)$ transition) taken on different days agree within 500~kHz, showing that the consistency of the positions of the rotational levels of \api\ is sub-MHz. The relatively large systematic uncertainty due to mode pulling and chirp will only enter in the value of the band origin. The root-mean-square deviation of 14 measurements of the $^{12}$C$^{16}$O $R_{2}(0)$ transition is about 2~MHz. We have set the uncertainty of the $R_{2}(0)$ transition, and hence the systematic uncertainty on all transitions, conservatively at 5~MHz. Note that Salumbides \emph{et al.}~\cite{Salumbides} have circumvented this systematic offset of the absolute calibration by putting a small pinhole in the pulsed beam, therewith selecting a portion of the laser beam with smaller wave-front distortions. We could not apply this method here, as it resulted in a too large decrease in signal.

\begin{table}
\begin{ruledtabular}\begin{tabular}{lc}
Source&Uncertainty(MHz)\\
\hline
\bigstrut
Relative:&\\
~~~~Zeeman effect&$<$0.1\\
~~~~Stark effect& $<$0.03\\
~~~~AC-Stark effect&$<$0.1\\
~~~~Comb absolute frequency determination&$\ll$0.15\\
~~~~Doppler effect &$<$0.2\\
~~~~Peak determination&0.3\\
~~~~CW-pulse offset&0.2\\
Absolute:&\\
~~~~CW-pulse offset&5\\
\end{tabular}\end{ruledtabular}
\caption{The uncertainty budget of the measured transitions.}
\label{tab:error}
\end{table}

\section{Analysis}
\label{sec:ana}

\subsection{Effective Hamiltonian and least square fitting for $^{12}$C$^{16}$O and $^{13}$C$^{16}$O}
\label{subsec:C12O16}

\begin{table*}
\begin{ruledtabular}\begin{tabular}{l|l}
$\langle ^3\Pi_0 | H | ^3\Pi_0 \rangle$&$E_\Pi+B(x+1)-D(x^2+4x+1)-A-2A_J(x+1)-C\mp C_{\delta}-\gamma+xB_0+(1\mp1)(4B_1^+-2B_0^+)$\\
$\langle ^3\Pi_1 | H | ^3\Pi_1 \rangle$&$E_\Pi+B(x+1)-D(x^2+6x-3)+2C-2\gamma-2B_0^++4B_1^+ $\\
$\langle ^3\Pi_2 | H | ^3\Pi_2 \rangle$&$E_\Pi+B(x-3)-D(x^2-4x+5)+A+2A_J(x-3)-C-\gamma+B_0^+(x-2)$\\
$\langle ^3\Pi_0 | H | ^3\Pi_1 \rangle$&$-\sqrt{2x}\left[ B-2D(x+1)-A_J-0.5\gamma+(1\mp2)B_1^+ \right] $ \\
$\langle ^3\Pi_0 | H | ^3\Pi_2 \rangle$&$ -\sqrt{x(x-2)}\left[ 2D \pm B_0^+\right] $ \\
$\langle ^3\Pi_1 | H | ^3\Pi_2 \rangle$&$-\sqrt{2(x-2)}\left[ B-2D(x-1)+A_J-0.5\gamma+B_1^+\right] $\\
\end{tabular}\end{ruledtabular}
\caption{The matrix form of the Hamiltonian used in the comprehensive fit. $x=J(J+1)$. Upper sign choice refers to \textit{e} levels, lower sign choice to \textit{f} levels.}
\label{tab:matrix}
\end{table*}

\begin{table}
\begin{ruledtabular}\begin{tabular}{l|l}
Carballo \emph{et al.} & Brown and Merer\\
\hline
\bigstrut
$E_\Pi$&$T+B-D-q/2$\\
$B$&$B-2D-q/2$\\
$D$&$D$\\
$A$&$A+A_D+\gamma+p/2$\\
$A_j$&$A_D/2$\\
$C$&$-2/3\lambda$\\
$C_{\delta}$&$o$\\
$\gamma$&$\gamma+p/2$\\
$B_0^+$&$q/2$\\
$B_1^+$&$p/4+q/2$\\
\end{tabular}\end{ruledtabular}
\caption{The conversion factors between the molecular constants used in the effective Hamiltonian of Carballo \emph{et al.}~\cite{Carballo} and Brown and Merer~\cite{BrownMerer}.}
\label{tab:rosetta}
\end{table}

The effective Hamiltonian for a $^3 \Pi$ state has been derived by several authors~\cite{Field,BrownMerer}. We have used 
the effective Hamiltonian from Field \emph{et al.}~\cite{Field} with the additions and corrections discussed by Carballo \emph{et al.}~\cite{Carballo}. The matrix elements for this effective Hamiltonian are listed in Table~\ref{tab:matrix}. As discussed by Carballo \emph{et al.}~\cite{Carballo}, this Hamiltonian is equivalent to the effective Hamiltonian derived by Brown and Merer~\cite{BrownMerer}, but the molecular constants used in these Hamiltonians have a slightly different physical meaning, which will have consequences for the mass scaling discussed in Sec.~\ref{subsec:Isotopes}. In Table~\ref{tab:rosetta}, relations between the constants in the Hamiltonian of Brown and Merer and those in the Hamiltonian of Field \emph{et al.} are listed for clarity.

A least-squares fitting routine was written in Mathematica to obtain the molecular constants of the effective Hamiltonian. We have verified that our fitting routine exactly reproduces the results of Carballo \emph{et al.}~\cite{Carballo} and that it is consistent with PGopher~\cite{pgopher}. For $^{12}$C$^{16}$O, we have fitted our optical data simultaneously with lambda-doubling transition frequencies in the rf domain measured by Wicke \emph{et al.}~\cite{Wicke:1972} and the rotational transition frequencies in the mw domain measured by Carballo \emph{et al.}~\cite{Carballo} and Wada and Kanamori~\cite{Wada}.
The fitted set consists of 9~rf transitions, 28~mw transitions and 14~optical transitions. The different data sets were given a weight of one over the square of the measurement uncertainties, taken as 50~kHz for the mw and rf data and 1~MHz for the optical data. The molecular constants for the \X\ state of $^{12}$C$^{16}$O are taken from Winnewisser \emph{et al.}~\cite{Winnewisser:1997}. As discussed by Carballo \emph{et al.}~\cite{Carballo}, $\gamma$ and $A_J$ can not be determined simultaneously from data of a single isotopologue, therefore $\gamma$ was fixed to zero. The first column of Table~\ref{tab:fitpars} lists the different constants obtained from our fit for $^{12}$C$^{16}$O. The deviations between the observed and fitted transition frequencies are listed in Table~\ref{tab:fitresults}.

\begin{table*}
\begin{ruledtabular}
\begin{tabular}{lrrrrr}
Molecular& $K_{\mu}^X$&$^{12}$C$^{16}$O&$^{13}$C$^{16}$O&$^{13}$C$^{16}$O&$^{13}$C$^{18}$O\\
Constant&&Fitted&Fitted&Scaled&Scaled\\
\hline
\bigstrut
$E$&0 &1453190243.1(8)	&	1453340486.4(7)	&	1453340489.2(9)	&	1453500584(3)	\\
$B$&-1	&50414.24(3)	&	48198.28(7)	&	48197.73	&	45797.49	\\
$D$&-2	&0.1919(3)	&	0.1861(13)	&	0.1753	&	0.1582	\\
$A$&0	&1242751.3(10)	&	1242807.6(10)	&	1242806.1	&	1242866.7	\\
$A_j$&-1	&-5.732(8)	&	-5.51(2)	&	-5.479	&	-5.206	\\
$C$&0	&-538.4(6)	&	-536.8(3)	&	-538.8	&	-539.5	\\
$C_{\delta}$&0	&26040.4(14)	&	26042.6(11)	&	26044.6	&	26049.4	\\
$\gamma$&-1&0&0&0&0\\
$B_0^+$&-2	&0.841(9)	&	0.775(7)	&	0.768	&	0.693	\\
$B_1^+$&-1	&39.67(5)	&	37.81(4)	&	37.94	&	36.08	\\
$a$&---	&---	&	162.2(3)	&	161.9(10)	&	161.9	\\
$b$&---	&---	&	638.0(6)	&	638(2)	&	638	\\
$c$&---	&---	&	8.3(3)	&	8.5(10)	&	8.5	\\
$d$&---	&---	&	107.3(10)	&	105(3)	&	105	\\
\end{tabular}
\caption{Molecular constants for the \api\ state of $^{12}$C$^{16}$O and $^{12}$C$^{16}$O, obtained from a simultaneous fit to rf, mw and optical data. $\gamma$ is fixed at zero in the analysis. The one standard deviation uncertainty from the fits is listed after each value, in units of the last digit. The values for the band origins have an additional uncertainty of 5~MHz due to the CW-pulse offset in the present optical measurements. For $^{13}$C$^{16}$O and  $^{13}$C$^{18}$O, the table also lists the values that were obtained by scaling the constants of $^{12}$C$^{16}$O. For $^{13}$C$^{16}$O, when the constants are scaled, the hyperfine constants are fitted together with the value of the band origin, while the scaled constants are kept fixed, resulting in slightly different values for these constants. For $^{13}$C$^{18}$O the hyperfine parameters of $^{13}$C$^{16}$O are used.}
\label{tab:fitpars}
\end{ruledtabular}
\end{table*}

All measured transitions could be fitted to approximately their respective uncertainties. However, the total root mean square (rms) of the residuals of the rf and mw data is significantly increased when the optical data is included; Carballo reported rms residuals of 27~kHz, while we find rms residuals of 52~kHz. The rms of the residuals of the fitted $^{12}$C$^{16}$O optical transitions is equal to 3.3~MHz. 

The molecular constants found from the fit to the combined data agree well with the constants found from a fit to the rf and mw data alone, but with largely decreased uncertainties. The uncertainty of $A$, $C$ and $C_{\delta}$, which are poorly constrained by the rotational and lambda-doubling transitions alone, are reduced by more than a factor of 10. Somewhat unexpectedly, the uncertainty of several other constants, including $B$, are also substantially decreased by the fit to the combined data. This can be understood from the fact that the mw data does not constrain $B$, but rather a combination of $A$ and $B$. This also explains why the uncertainty of $B$ as obtained from a fit to the rotational and lambda-doubling transitions is $\sim$300~kHz, whereas the rotational transitions have a quoted uncertainty of 5-8~kHz and are fitted with an rms uncertainty of 27~kHz~\cite{Carballo}. Our optical data directly probe $A$, and the more precise value of $A$ results in turn in a more precise value of $B$. Adding the optical data results in an uncertainty in $B$ of 30~kHz, much closer to the value one would expect from the precision of the recorded mw transitions. Altogether, the new set of constants is more balanced and adequately describes the \api\ state.

As seen from Table~\ref{tab:fitresults}, the residuals of the optical transitions probing the $\Omega=1$ manifold are larger than those to the $\Omega=0$ and $\Omega=2$ manifolds, which is surprising as these transitions are the strongest transitions in the spectra and are measured with a higher signal-to-noise ratio than the other transitions. We have investigated whether this might be explained by perturbations arising from the $a'^3\Sigma^+$ or $D^1\Delta$ state. When perturbations with the $a'^3\Sigma^+$ state were included, using the perturbation parameters from Carballo \emph{et al.}~\cite{Carballo}, the residuals decreased marginally. A slight improvement was obtained by including a perturbation with the $D^1\Delta$ state. However, as the perturbation parameters of the coupling between the \api\ and $D^1\Delta$ states are unknown, it is unclear if this improvement is genuine.

A similar analysis has been performed for $^{13}$C$^{16}$O. We have included the hyperfine interaction in the effective Hamiltonian following Brown \emph{et al.}~\cite{Brown:1977}. Only terms that are diagonal in $J$ were included, since contributions from off-diagonal terms are estimated to be smaller than 100 kHz~\cite{Gammon:1971}. We have fitted our optical data simultaneously with lambda-doublet transition frequencies measured by Gammon \emph{et al.}~\cite{Gammon:1971} and the rotational transition frequencies measured by Saykally \emph{et al.}~\cite{Saykally:1987}. The fitted set consisted of 19~rf transitions, 4~mw transitions and 24 optical transitions. The molecular constants for the \X\ are taken from Klapper \emph{et al.}~\cite{Klapper:2000}. The molecular constants resulting from the fit are listed in Table~\ref{tab:fitpars}. The difference between the observed transition frequencies and the frequencies from the fit are listed in Table~\ref{tab:fitresults}. The rms of the residuals of the fitted $^{13}$C$^{16}$O optical transitions is equal to 0.9~MHz. 

For $^{13}$C$^{18}$O, as for the other isotopologues, no previous measurements on the \api\ were found in the literature, except for four mw-transitions in $^{12}$C$^{18}$O~\cite{Saykally:1987}. Hence, no fit has been attempted. 

\subsection{Mass scaling}
\label{subsec:Isotopes}

An important motivation for this work was to validate the mass scaling of the energy levels of the \api\ state, and to confirm the sensitivity to a possible variation of the proton-to-electron mass ratio for a selected number of level spittings. In the literature, the reduced mass of the molecule is frequently denoted by the symbol $\mu$. In this paper, we will use $\mu$ to denote the proton-to-electron mass ratio, and will denote the reduced mass of the molecule by $\mu_{red}$. As we will see in Sec.~\ref{subsec:deltamu}, $\mu_{red}$ is linearly proportional to $\mu$ which is defined for the various isotopologues $^x$C$^y$O as 

\begin{equation}
\mu_{red}^{x,y}=\frac{m_C^x \cdot m_O^y}{m_C^x + m_O^y}.
\label{Eq:mured}
\end{equation}

\noindent 
The molecular constants as determined from the fits are effective molecular constants for the $v=0$ level of the \api\ state. In general, an effective molecular constant $X_{e,v}$ can be expressed as

\begin{multline}
X_{e,v} = X_{e} + \alpha_{X,1}(v+1/2) \\
+ \alpha_{X,2}(v+1/2)^2 + \ldots . 
\label{Eq:vserie}
\end{multline}

\noindent
Note that for the constants $A$ and $B$ the second term of Eq.~(\ref{Eq:vserie}) has a minus sign by convention~\cite{Havenith}. Thus, the mass dependence of every constant consists of the mass dependence of $X_{e}$ and a correction due to the vibrational dependence of $X_{e,v}$. The second column of Table~\ref{tab:fitpars} lists the dependence of the molecular constants, $X_{e}$, on the reduced mass of the molecule, $\mu_{red}$. The effective molecular constant, $X'_{e,v}$, of an isotopologue with a reduced mass $\mu'_{red}$ then becomes

\begin{equation}
\begin{split}
X'_{e,v} = &\left(\frac{\mu'_{red}}{\mu_{red}}\right)^{K^{X}_{\mu}}X_{e}\\
 + &\left(\frac{\mu'_{red}}{\mu_{red}}\right)^{K^{X}_{\mu}+\frac{1}{2}}
\alpha_{X,1}(v+1/2) \\
+&\left(\frac{\mu'_{red}}{\mu_{red}}\right)^{K^{X}_{\mu}+1}
\alpha_{X,2}(v+1/2)^2 + \ldots ,
\end{split}
\label{Eq:scaling}
\end{equation}

\noindent
where $X_{e}$, $\alpha_{X,1}$ and $\alpha_{X,2}$ are the constants for the isotopologue with reduced mass $\mu_{red}$ and

\begin{equation}
K^{X}_{\mu} = \frac{\mu}{X_{e}}\frac{\partial X_{e}}{\partial \mu}.
\label{Eq:KXmu}
\end{equation} 

\noindent
Note that we use $\mu$ in Eq.~(\ref{Eq:KXmu}) rather than $\mu_{red}$, see below.

As we have only measured transitions in the $v=0$ band, the vibrational dependencies of the molecular constants cannot be extracted from our data. Hence, we have used the ratios between $X_{e}$, $\alpha_{X,1}$ and $\alpha_{X,2}$ determined by Havenith \emph{et al.}~\cite{Havenith} to scale our constants. The molecular constants for $^{13}$C$^{16}$O and $^{13}$C$^{18}$O, found by scaling the constants of $^{12}$C$^{16}$O via the outlined procedure, are listed in the third and fourth column of Table~\ref{tab:fitpars}, respectively. For $^{13}$C$^{16}$O, the value of the band origin and the hyperfine constants were determined by fitting the data while the other constants were fixed at the scaled values. For $^{13}$C$^{18}$O, the hyperfine constants were taken to be identical to those of $^{13}$C$^{16}$O. The data for the \X\ state was calculated from Puzzarini \emph{et al.}~\cite{Puzzarini:2003} and only the value of the band origin was fitted. The differences between the observed transition frequencies and the frequencies calculated using the scaled molecular constants are listed in Table~\ref{tab:fitresults}. As is seen the correspondence is satisfactory. The rms of the residuals of the $^{13}$C$^{16}$O and $^{13}$C$^{18}$O data with the frequencies found by scaling the molecular constants is equal to 5.1~MHz and 8.3~MHz, respectively.

\begin{figure}
\begin{center}
\includegraphics[width=1.1\linewidth]{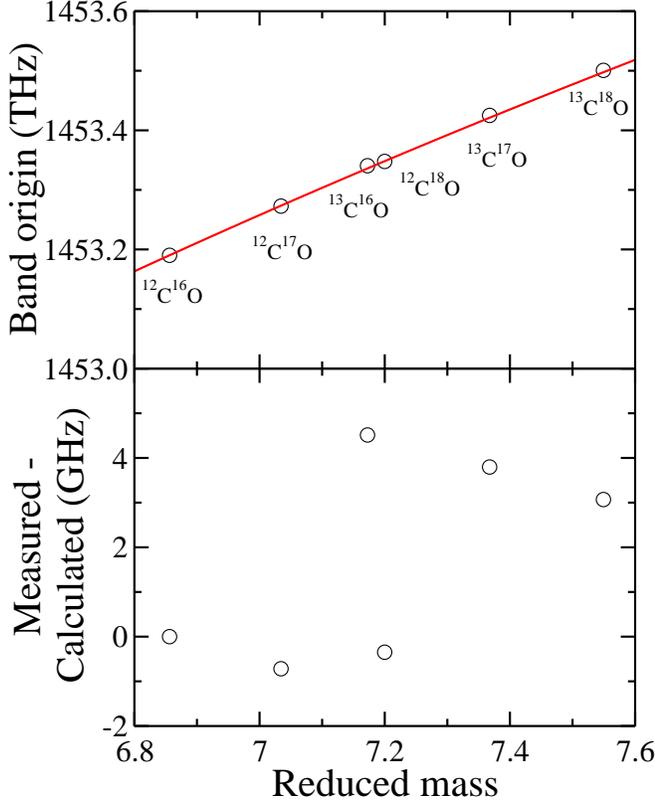}
\end{center}
\caption{(Color online) In the upper panel, the measured values of the band origins of the \api\ state of the six stable isotopologues of CO are plotted as a function of the reduced mass. The solid line shows the value of the band origin scaled with respect to $^{12}$C$^{16}$O. In the lower panel the difference between the measured and calculated values of the band origins are plotted.}
\label{fig:isotopeshift}
\end{figure}

Until now the value of the band origin, $E_\Pi$, was treated as a free parameter without considering its proper mass scaling. The value of the band origin consists of: (i) A pure electronic part, that scales as $(\mu_{red})^0$ except for a small correction due to the finite mass of the nuclei, known as the normal mass shift, or Bohr-shift, which is proportional to the reduced mass of the nuclei-electron system. (ii) A vibronic part that can be expanded in a power series of $(v+1/2)$ and contains the difference in zero-point energies in the \api\ and \X\ states. (iii) A rotational part, equal to $B - D$, that was absorbed in the value of the band origin in our definition of the effective Hamiltonian. (iv) The specific mass shift, dependent on the electron correlation function. (v) Nuclear-size effects, dependent on the probability density function of the electrons at the nucleus.

In the upper panel of Fig.~\ref{fig:isotopeshift}, the derived values of the band origin of the \api\ state of the six stable isotopologues of CO are plotted as a function of the reduced mass. The solid line shows how the value of the band origin scales when effects (i-iii), which are expected to be dominant, are included. The used formulas for the normal mass shift, the vibrational and the rotational parts are: 

\begin{equation}
\begin{split}
\Delta E_\Pi=\Delta E_{nms}+\Delta E_{vib}+\Delta E_{rot}
\end{split}
\label{Eq:Escaling}
\end{equation}

\noindent with

\begin{equation}
\begin{split}
\Delta E_{nms}=E_{\Pi \mathrm{0}}\left[\left(\frac{\mu'_{red} \cdot m_{el}}{\mu'_{red} + m_{el}}\right)/\left(\frac{\mu_{red} \cdot m_{el}}{\mu_{red} + m_{el}}\right)-1\right],
\end{split}
\label{Eq:Escaling1}
\end{equation} 

\begin{equation}
\begin{split}
\Delta E_{vib}&= \frac{1}{2} \left(\omega _{e\Pi}- \omega _{e\Sigma}\right)\left(\left(\frac{\mu_{red}}{\mu'_{red}}\right)^{1/2}-1\right)\\
-&\frac{1}{4} \left(\omega _{e\Pi}x_{e\Pi}- \omega _{e\Sigma}x_{e\Sigma}\right) \left(\frac{\mu_{red}}{\mu'_{red}}-1\right)\\
+&\frac{1}{8} \left(\omega _{e\Pi}y_{e\Pi}- \omega _{e\Sigma}y_{e\Sigma}\right) \left(\left(\frac{\mu_{red}}{\mu'_{red}}\right)^{3/2}-1\right),
\end{split}
\label{Eq:Escaling2}
\end{equation} 

\begin{equation}
\begin{split}
\Delta E_{rot}=B_{\mathrm{0}}\left(\frac{\mu_{red}}{\mu'_{red}}-1\right)-D_{\mathrm{0}}\left(\left(\frac{\mu_{red}}
{\mu'_{red}}\right)^2-1\right)
\end{split}
\label{Eq:Escaling3}
\end{equation}

\noindent where $\Delta E_\Pi$ is the shift in the value of the band origin as a function of the reduced mass, $\mu'_{red}$, with respect to a given isotopologue with reduced mass $\mu_{red}$, band origin $E_{\Pi \mathrm{0}}$ and rotational constants $B_{\mathrm{0}}$ and $D_{\mathrm{0}}$. $m_{el}$ is the mass of the electron. The shift was calculated with respect to $^{12}$C$^{16}$O. For the \X\ state the vibrational constants were obtained from fitting to the data from Coxon \emph{et al.}~\cite{Coxon}, while for the \api\ state the constants from Havenith \emph{et al.}~\cite{Havenith} were used. In the lower panel, the difference between the experimental and calculated values of the band origins for the different isotopologues are plotted. The experimental and scaled values of the band origins deviate by a few GHz, which is more than expected given the precision of the data used in this analysis. Most surprising is the difference between the values of the band origins of $^{13}$C$^{16}$O and $^{12}$C$^{18}$O, which have nearly equal reduced masses (7.18 vs. 7.21 amu, respectively). This suggests a breakdown of the Born-Oppenheimer approximation. 

We have re-analyzed experimental data pertaining to all six stable isotopologues of CO for the E$^1\Pi$ ($v=1$) state~\cite{Ubachs:2000} and the C$^1\Sigma^+$ ($v=1$) state~\cite{Ubachs:2001} and found a similar deviation as found for the \api, both in size and in direction. This suggests that the ground state is probably the source of the discrepancy. Calculations using the LEVEL program~\cite{level} were performed to estimate the effects of the breakdown of the Born-Oppenheimer approximation in the ground-state of CO following the approach by Coxon and Hajigeorgiou~\cite{Coxon}.
The calculated energy difference between the $^{12}$C$^{18}$O and $^{13}$C$^{16}$O isotopologues was approximately 50 times smaller than observed in our measurements and is thus insufficient to explain the observed effect~\cite{LeRoyprivatecommunications}.

The specific mass shift was not included in our analysis. It is however proportional to $\mu_{red}$, and can thus not explain the observed large difference between $^{12}$C$^{18}$O and $^{13}$C$^{16}$O. Nuclear size effects could in principle cause a similar isotope shift as observed, but the difference in nuclear charge radius between $^{17}$O and $^{18}$O is approximately 10 times larger than the difference between $^{16}$O and $^{17}$O, whereas the observed difference between the values of the band origins in the isotopologues with these oxygen isotopes is similar in size~\cite{Angeli}. It is therefore unlikely that the observed isotopic effect is due to a nuclear-size effect. 

\subsection{CO(\api) as a target system for probing $\partial \mu / \partial t$.}
\label{subsec:deltamu}

\begin{table*}
\begin{ruledtabular}\begin{tabular}{cl D..{1} D..{1} D..{2}}
Isotopologue & Transition & \mathrm{Measured~(MHz)}& \mathrm{Meas} _\cdot \mathrm{-Calc}. \mathrm{(MHz)}& K_\mu\\
\hline
\bigstrut
$^{12}$C$^{16}$O & $J=6,~\Omega =1,+ \rightarrow J=4,~\Omega=2,+$ &19270.1 & 3.5 & 27.8\\
~ & $J=6,~\Omega =1,- \rightarrow J=4,~\Omega=2,-$ & 16057.7 & 4.7 & 33.7\\
~ & $J=6,~\Omega =1,+ \rightarrow J=8,~\Omega=0,+$ & -1628.3 & -3.3 & -334\\
~ & $J=6,~\Omega =1,- \rightarrow J=8,~\Omega=0,-$ & -19406.7 & 4.5 & -27.3\\
$^{13}$C$^{16}$O & $J=6,~\Omega =1,+,F=3.5 \rightarrow J=4,~\Omega=2,+,F=6.5$ & 43005.8 & 0.1 &12.9\\
~ & $J=6,~\Omega =1,-,F=3.5 \rightarrow J=4,~\Omega=2,-,F=6.5$ & 39988.0 & 0.7 &12.2\\
~ & $J=6,~\Omega =1,+,F=5.5 \rightarrow J=8,~\Omega=0,+,F=8.5$ & 22329.6 & 6.0 &23.5\\
~ & $J=6,~\Omega =1,-,F=5.5 \rightarrow J=8,~\Omega=0,-,F=8.5$ & 4003.4 & 5.0 &128\\
$^{13}$C$^{18}$O & $J=6,~\Omega =1,+,F=3.5 \rightarrow J=4,~\Omega=2,+,F=6.5$ & 69062.0 & -7.1 &7.22\\
~ & $J=6,~\Omega =1,-,F=3.5 \rightarrow J=4,~\Omega=2,-,F=6.5$ & 66277.5 & -3.6 &7.60\\
\end{tabular}\end{ruledtabular}
\caption{Measured level splittings between near-degenerate levels in CO (\api); frequencies are listed in terms of two-photon microwave transitions bridging the intervals. The Meas.-Calc. column lists the difference between the measured frequency and the frequency calculated using the constants listed in Table~\ref{tab:fitpars}. The sensitivities to a possible variation of the proton-to-electron mass ratio are listed in the last column.}
\label{tab:ndg}
\end{table*}

\begin{figure}
\begin{center}
\includegraphics[width=1.1\linewidth]{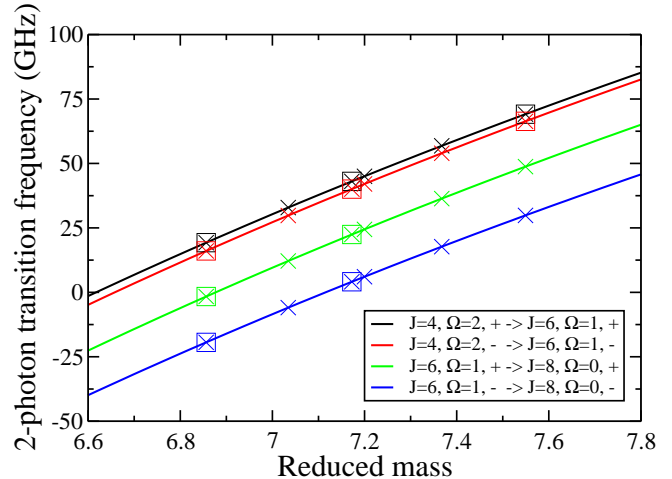}
\end{center}
\caption{(Color online) The frequencies of two-photon mw transitions between four near degeneracies as a function the reduced mass. The solid lines show the values obtained from mass scaling the molecular constants of $^{12}$C$^{16}$O. The crosses indicate the calculated frequencies of the transitions in the six stable isotopologues, whereas the boxes show the values obtained from differences between measured frequencies. The four different transitions are listed in the legend as they appear in the graph, from top to bottom.}
\label{fig:ndg}
\end{figure}

In Fig.~\ref{fig:ndg}, the energies of the two-photon mw transitions between the near degenerate levels of the different $\Omega$-manifolds are shown as a function of $\mu_{red}$. The solid lines show the values obtained from scaling the molecular constants of $^{12}$C$^{16}$O using the mass-scaling relations discussed in Sec.\ref{subsec:Isotopes}. The crosses indicate the reduced masses of the six stable isotopologues, whereas the boxes show the values directly obtained from the measurements. The strong energy dependence on the reduced mass is indicative of a high sensitivity to the proton-to-electron mass ratio. The measured and calculated values for the two-photon microwave transitions between the nearly-degenerate levels are listed in Table~\ref{tab:ndg}. Our measurements reduce the uncertainties of the two-photon transitions to $\approx$1~MHz, a factor 20 better than before.

The mass of the proton is much larger than the masses of the constituent quarks, consequently, the mass of the proton is related to the strength of the forces between the quarks; $\Lambda_{QCD}$, the scale of quantum chromodynamics~\cite{Flambaum:2004}. As the same argument holds for the neutron-to-electron mass ratio, a possible variation of the proton-to-electron mass ratio is expected to be accompanied by a similar variation of the neutron-to-electron mass ratio~\cite{Dent:2007}. With this assumption, it follows that

\begin{equation}
K_{\mu} = \frac{\mu}{\nu}\frac{\partial \nu}{\partial \mu}=
\frac{\mu_{red}}{\nu}\frac{\partial \nu}{\partial \mu_{red}}.
\label{Eq:trivial}
\end{equation} 

The sensitivity of a transition to a possible variation of the proton-to-electron mass ratio $\mu$ can now be calculated using the mass-scaling relations discussed before. The sensitivities for the two-photon microwave transitions are listed in the last column of Table~\ref{tab:ndg}. These coefficients have been calculated with an accuracy of 0.3-0.05\%.

It is instructive to compare the sensitivity to what one would expect in a pure Hund's case (a). For the transition at 1628.3~MHz we expect a sensitivity of $K_\mu=A/2\nu=1242751.3/2\times1628.3\approx382$ which is 12\%  larger than calculated using the model that includes the coupling between the different $\Omega$-manifolds~\footnote{Note that Bethlem and Ubachs~\cite{Bethlem:2009} erroneously used $K_\mu=A/\nu$ instead of $K_\mu=A/2\nu$.}.

\section{Conclusion}

UV frequency metrology has been performed on the \api\ - \X\ (0,0) band of various isotopologues of CO using a frequency-quadrupled injection-seeded narrow-band pulsed Ti:Sa laser referenced to a frequency comb laser.

We have fitted our optical data for $^{12}$C$^{16}$O together with the lambda-doubling transitions of Wicke \emph{et al.}~\cite{Wicke:1972} and the rotational transitions of Carballo \emph{et al.}~\cite{Carballo} and Wada and Kanamori~\cite{Wada}. Adding the optical data resulted in a large decrease of the uncertainties of $A$, $C$ and $C_{\delta}$, and a smaller decrease in uncertainty in the other constants.

From our measurements we obtain the value of the band origin with an uncertainty of 5~MHz, a 30-fold improvement compared to the value obtained from absorption measurements by Field \emph{et al.}~\cite{Field}. We have also measured the value of the band origin in different isotopes and found an unexpected behavior of the isotope shifts, probably due to a breakdown of the Born-Oppenheimer approximation. 

Our main motivation for this study was to obtain more accurate values for the 2-photon transitions between near degenerate rotational levels in different $\Omega$-manifolds and validate the large sensitivity coefficients predicted for these transitions. The calculated values of the transitions agree to within a few MHz with the measured values, giving confidence in the calculated values of $K_\mu$.

\section{Acknowledgements}
We thank Robert J. Le Roy for his calculations on breakdown of the Born-Oppenheimer approximation in the ground state of CO and Gerrit Groenenboom for his calculations on the lifetime of the $J=0,\,\Omega=0$ level. This work is financially supported by the Netherlands Foundation for Fundamental Research of Matter (FOM) (project 10PR2793 and program ``Broken mirrors and drifting constants''). H.L.B. acknowledges financial support from NWO via a VIDI-grant, and from the ERC via a Starting Grant.


\begin{thebibliography}{39}%
\makeatletter
\providecommand \@ifxundefined [1]{%
 \@ifx{#1\undefined}
}%
\providecommand \@ifnum [1]{%
 \ifnum #1\expandafter \@firstoftwo
 \else \expandafter \@secondoftwo
 \fi
}%
\providecommand \@ifx [1]{%
 \ifx #1\expandafter \@firstoftwo
 \else \expandafter \@secondoftwo
 \fi
}%
\providecommand \natexlab [1]{#1}%
\providecommand \enquote  [1]{``#1''}%
\providecommand \bibnamefont  [1]{#1}%
\providecommand \bibfnamefont [1]{#1}%
\providecommand \citenamefont [1]{#1}%
\providecommand \href@noop [0]{\@secondoftwo}%
\providecommand \href [0]{\begingroup \@sanitize@url \@href}%
\providecommand \@href[1]{\@@startlink{#1}\@@href}%
\providecommand \@@href[1]{\endgroup#1\@@endlink}%
\providecommand \@sanitize@url [0]{\catcode `\\12\catcode `\$12\catcode
  `\&12\catcode `\#12\catcode `\^12\catcode `\_12\catcode `\%12\relax}%
\providecommand \@@startlink[1]{}%
\providecommand \@@endlink[0]{}%
\providecommand \url  [0]{\begingroup\@sanitize@url \@url }%
\providecommand \@url [1]{\endgroup\@href {#1}{\urlprefix }}%
\providecommand \urlprefix  [0]{URL }%
\providecommand \Eprint [0]{\href }%
\providecommand \doibase [0]{http://dx.doi.org/}%
\providecommand \selectlanguage [0]{\@gobble}%
\providecommand \bibinfo  [0]{\@secondoftwo}%
\providecommand \bibfield  [0]{\@secondoftwo}%
\providecommand \translation [1]{[#1]}%
\providecommand \BibitemOpen [0]{}%
\providecommand \bibitemStop [0]{}%
\providecommand \bibitemNoStop [0]{.\EOS\space}%
\providecommand \EOS [0]{\spacefactor3000\relax}%
\providecommand \BibitemShut  [1]{\csname bibitem#1\endcsname}%
\let\auto@bib@innerbib\@empty
\bibitem [{\citenamefont {Cameron}(1926)}]{Cameron}%
  \BibitemOpen
  \bibfield  {author} {\bibinfo {author} {\bibfnamefont {W.~H.~B.}\
  \bibnamefont {Cameron}},\ }\href@noop {} {\bibfield  {journal} {\bibinfo
  {journal} {Philos. Mag.}\ }\textbf {\bibinfo {volume} {1}},\ \bibinfo {pages}
  {405} (\bibinfo {year} {1926})}\BibitemShut {NoStop}%
\bibitem [{\citenamefont {Freund}\ and\ \citenamefont
  {Klemperer}(1965)}]{Freund}%
  \BibitemOpen
  \bibfield  {author} {\bibinfo {author} {\bibfnamefont {R.~S.}\ \bibnamefont
  {Freund}}\ and\ \bibinfo {author} {\bibfnamefont {W.}~\bibnamefont
  {Klemperer}},\ }\href@noop {} {\bibfield  {journal} {\bibinfo  {journal} {J.
  Chem. Phys.}\ }\textbf {\bibinfo {volume} {43}},\ \bibinfo {pages} {2422}
  (\bibinfo {year} {1965})}\BibitemShut {NoStop}%
\bibitem [{\citenamefont {Wicke}\ \emph {et~al.}(1972)\citenamefont {Wicke},
  \citenamefont {Field},\ and\ \citenamefont {Klemperer}}]{Wicke:1972}%
  \BibitemOpen
  \bibfield  {author} {\bibinfo {author} {\bibfnamefont {B.~G.}\ \bibnamefont
  {Wicke}}, \bibinfo {author} {\bibfnamefont {R.~W.}\ \bibnamefont {Field}}, \
  and\ \bibinfo {author} {\bibfnamefont {W.}~\bibnamefont {Klemperer}},\
  }\href@noop {} {\bibfield  {journal} {\bibinfo  {journal} {J. Chem. Phys.}\
  }\textbf {\bibinfo {volume} {56}},\ \bibinfo {pages} {5758} (\bibinfo {year}
  {1972})}\BibitemShut {NoStop}%
\bibitem [{\citenamefont {Saykally}\ \emph {et~al.}(1987)\citenamefont
  {Saykally}, \citenamefont {Dixon}, \citenamefont {Anderson}, \citenamefont
  {Szanto},\ and\ \citenamefont {Woods}}]{Saykally:1987}%
  \BibitemOpen
  \bibfield  {author} {\bibinfo {author} {\bibfnamefont {R.~J.}\ \bibnamefont
  {Saykally}}, \bibinfo {author} {\bibfnamefont {T.~A.}\ \bibnamefont {Dixon}},
  \bibinfo {author} {\bibfnamefont {T.~G.}\ \bibnamefont {Anderson}}, \bibinfo
  {author} {\bibfnamefont {P.~G.}\ \bibnamefont {Szanto}}, \ and\ \bibinfo
  {author} {\bibfnamefont {R.~C.}\ \bibnamefont {Woods}},\ }\href@noop {}
  {\bibfield  {journal} {\bibinfo  {journal} {J. Chem. Phys.}\ }\textbf
  {\bibinfo {volume} {87}},\ \bibinfo {pages} {6423} (\bibinfo {year}
  {1987})}\BibitemShut {NoStop}%
\bibitem [{\citenamefont {Carballo}\ \emph {et~al.}(1988)\citenamefont
  {Carballo}, \citenamefont {Warner}, \citenamefont {Gudeman},\ and\
  \citenamefont {Woods}}]{Carballo}%
  \BibitemOpen
  \bibfield  {author} {\bibinfo {author} {\bibfnamefont {N.}~\bibnamefont
  {Carballo}}, \bibinfo {author} {\bibfnamefont {H.~E.}\ \bibnamefont
  {Warner}}, \bibinfo {author} {\bibfnamefont {C.~S.}\ \bibnamefont {Gudeman}},
  \ and\ \bibinfo {author} {\bibfnamefont {R.~C.}\ \bibnamefont {Woods}},\
  }\href@noop {} {\bibfield  {journal} {\bibinfo  {journal} {J. Chem. Phys.}\
  }\textbf {\bibinfo {volume} {88}},\ \bibinfo {pages} {7273} (\bibinfo {year}
  {1988})}\BibitemShut {NoStop}%
\bibitem [{\citenamefont {Wada}\ and\ \citenamefont {Kanamori}(2000)}]{Wada}%
  \BibitemOpen
  \bibfield  {author} {\bibinfo {author} {\bibfnamefont {A.}~\bibnamefont
  {Wada}}\ and\ \bibinfo {author} {\bibfnamefont {H.}~\bibnamefont
  {Kanamori}},\ }\href@noop {} {\bibfield  {journal} {\bibinfo  {journal} {J.
  Mol. Spectrosc.}\ }\textbf {\bibinfo {volume} {200}},\ \bibinfo {pages} {196}
  (\bibinfo {year} {2000})}\BibitemShut {NoStop}%
\bibitem [{\citenamefont {Havenith}\ \emph {et~al.}(1988)\citenamefont
  {Havenith}, \citenamefont {Bohle}, \citenamefont {Werner},\ and\
  \citenamefont {Urban}}]{Havenith}%
  \BibitemOpen
  \bibfield  {author} {\bibinfo {author} {\bibfnamefont {M.}~\bibnamefont
  {Havenith}}, \bibinfo {author} {\bibfnamefont {W.}~\bibnamefont {Bohle}},
  \bibinfo {author} {\bibfnamefont {J.}~\bibnamefont {Werner}}, \ and\ \bibinfo
  {author} {\bibfnamefont {W.}~\bibnamefont {Urban}},\ }\href@noop {}
  {\bibfield  {journal} {\bibinfo  {journal} {Molec. Phys.}\ }\textbf {\bibinfo
  {volume} {64}},\ \bibinfo {pages} {1073} (\bibinfo {year}
  {1988})}\BibitemShut {NoStop}%
\bibitem [{\citenamefont {Davies}\ and\ \citenamefont {Martin}(1987)}]{Davies}%
  \BibitemOpen
  \bibfield  {author} {\bibinfo {author} {\bibfnamefont {P.~B.}\ \bibnamefont
  {Davies}}\ and\ \bibinfo {author} {\bibfnamefont {P.~A.}\ \bibnamefont
  {Martin}},\ }\href@noop {} {\bibfield  {journal} {\bibinfo  {journal} {Chem.
  Phys. Lett.}\ }\textbf {\bibinfo {volume} {136}},\ \bibinfo {pages} {527}
  (\bibinfo {year} {1987})}\BibitemShut {NoStop}%
\bibitem [{\citenamefont {Effantin}\ \emph {et~al.}(1982)\citenamefont
  {Effantin}, \citenamefont {Michaud}, \citenamefont {Roux}, \citenamefont
  {d'Incan},\ and\ \citenamefont {Verges}}]{Effantin:1982}%
  \BibitemOpen
  \bibfield  {author} {\bibinfo {author} {\bibfnamefont {C.}~\bibnamefont
  {Effantin}}, \bibinfo {author} {\bibfnamefont {F.}~\bibnamefont {Michaud}},
  \bibinfo {author} {\bibfnamefont {F.}~\bibnamefont {Roux}}, \bibinfo {author}
  {\bibfnamefont {J.}~\bibnamefont {d'Incan}}, \ and\ \bibinfo {author}
  {\bibfnamefont {J.}~\bibnamefont {Verges}},\ }\href@noop {} {\bibfield
  {journal} {\bibinfo  {journal} {J. Mol. Spectrosc.}\ }\textbf {\bibinfo
  {volume} {92}},\ \bibinfo {pages} {349} (\bibinfo {year} {1982})}\BibitemShut
  {NoStop}%
\bibitem [{\citenamefont {Field}\ \emph {et~al.}(1972)\citenamefont {Field},
  \citenamefont {Tilford}, \citenamefont {Howard},\ and\ \citenamefont
  {Simmons}}]{Field}%
  \BibitemOpen
  \bibfield  {author} {\bibinfo {author} {\bibfnamefont {R.~W.}\ \bibnamefont
  {Field}}, \bibinfo {author} {\bibfnamefont {S.~G.}\ \bibnamefont {Tilford}},
  \bibinfo {author} {\bibfnamefont {R.~A.}\ \bibnamefont {Howard}}, \ and\
  \bibinfo {author} {\bibfnamefont {J.~D.}\ \bibnamefont {Simmons}},\
  }\href@noop {} {\bibfield  {journal} {\bibinfo  {journal} {J. Mol.
  Spectrosc.}\ }\textbf {\bibinfo {volume} {44}},\ \bibinfo {pages} {347}
  (\bibinfo {year} {1972})}\BibitemShut {NoStop}%
\bibitem [{\citenamefont {Gammon}\ \emph {et~al.}(1971)\citenamefont {Gammon},
  \citenamefont {Stern}, \citenamefont {Lesk}, \citenamefont {Wicke},\ and\
  \citenamefont {Klemperer}}]{Gammon:1971}%
  \BibitemOpen
  \bibfield  {author} {\bibinfo {author} {\bibfnamefont {R.~H.}\ \bibnamefont
  {Gammon}}, \bibinfo {author} {\bibfnamefont {R.~C.}\ \bibnamefont {Stern}},
  \bibinfo {author} {\bibfnamefont {M.~E.}\ \bibnamefont {Lesk}}, \bibinfo
  {author} {\bibfnamefont {B.~G.}\ \bibnamefont {Wicke}}, \ and\ \bibinfo
  {author} {\bibfnamefont {W.}~\bibnamefont {Klemperer}},\ }\href@noop {}
  {\bibfield  {journal} {\bibinfo  {journal} {J. Chem. Phys.}\ }\textbf
  {\bibinfo {volume} {54}},\ \bibinfo {pages} {2136} (\bibinfo {year}
  {1971})}\BibitemShut {NoStop}%
\bibitem [{\citenamefont {Bethlem}\ and\ \citenamefont
  {Ubachs}(2009)}]{Bethlem:2009}%
  \BibitemOpen
  \bibfield  {author} {\bibinfo {author} {\bibfnamefont {H.~L.}\ \bibnamefont
  {Bethlem}}\ and\ \bibinfo {author} {\bibfnamefont {W.}~\bibnamefont
  {Ubachs}},\ }\href@noop {} {\bibfield  {journal} {\bibinfo  {journal}
  {Faraday Discuss.}\ }\textbf {\bibinfo {volume} {142}},\ \bibinfo {pages}
  {25} (\bibinfo {year} {2009})}\BibitemShut {NoStop}%
\bibitem [{\citenamefont {Uzan}(2003)}]{Uzan}%
  \BibitemOpen
  \bibfield  {author} {\bibinfo {author} {\bibfnamefont {J.~P.}\ \bibnamefont
  {Uzan}},\ }\href@noop {} {\bibfield  {journal} {\bibinfo  {journal} {Rev.
  Mod. Phys.}\ }\textbf {\bibinfo {volume} {75}},\ \bibinfo {pages} {403}
  (\bibinfo {year} {2003})}\BibitemShut {NoStop}%
\bibitem [{\citenamefont {Ubachs}\ \emph {et~al.}(2006)\citenamefont {Ubachs},
  \citenamefont {Buning}, \citenamefont {Eikema},\ and\ \citenamefont
  {Reinhold}}]{UbachsBuning}%
  \BibitemOpen
  \bibfield  {author} {\bibinfo {author} {\bibfnamefont {W.}~\bibnamefont
  {Ubachs}}, \bibinfo {author} {\bibfnamefont {R.}~\bibnamefont {Buning}},
  \bibinfo {author} {\bibfnamefont {K.~S.~E.}\ \bibnamefont {Eikema}}, \ and\
  \bibinfo {author} {\bibfnamefont {E.}~\bibnamefont {Reinhold}},\ }\href@noop
  {} {\bibfield  {journal} {\bibinfo  {journal} {J. Mol. Spectrosc.}\ }\textbf
  {\bibinfo {volume} {241}},\ \bibinfo {pages} {155} (\bibinfo {year}
  {2006})}\BibitemShut {NoStop}%
\bibitem [{\citenamefont {Gilijamse}\ \emph {et~al.}(2007)\citenamefont
  {Gilijamse}, \citenamefont {Hoekstra}, \citenamefont {Meek}, \citenamefont
  {Mets\"al\"a}, \citenamefont {{van de Meerakker}}, \citenamefont {Meijer},\
  and\ \citenamefont {Groenenboom}}]{Gilijamse}%
  \BibitemOpen
  \bibfield  {author} {\bibinfo {author} {\bibfnamefont {J.~J.}\ \bibnamefont
  {Gilijamse}}, \bibinfo {author} {\bibfnamefont {S.}~\bibnamefont {Hoekstra}},
  \bibinfo {author} {\bibfnamefont {S.~A.}\ \bibnamefont {Meek}}, \bibinfo
  {author} {\bibfnamefont {M.}~\bibnamefont {Mets\"al\"a}}, \bibinfo {author}
  {\bibfnamefont {S.~Y.~T.}\ \bibnamefont {{van de Meerakker}}}, \bibinfo
  {author} {\bibfnamefont {G.}~\bibnamefont {Meijer}}, \ and\ \bibinfo {author}
  {\bibfnamefont {G.~C.}\ \bibnamefont {Groenenboom}},\ }\href@noop {}
  {\bibfield  {journal} {\bibinfo  {journal} {J. Chem. Phys.}\ }\textbf
  {\bibinfo {volume} {127}},\ \bibinfo {pages} {221102} (\bibinfo {year}
  {2007})}\BibitemShut {NoStop}%
\bibitem [{\citenamefont {Klapper}\ \emph {et~al.}(2000)\citenamefont
  {Klapper}, \citenamefont {Lewen}, \citenamefont {Gendriesch}, \citenamefont
  {Belov},\ and\ \citenamefont {Winnewisser}}]{Klapper:2000}%
  \BibitemOpen
  \bibfield  {author} {\bibinfo {author} {\bibfnamefont {G.}~\bibnamefont
  {Klapper}}, \bibinfo {author} {\bibfnamefont {F.}~\bibnamefont {Lewen}},
  \bibinfo {author} {\bibfnamefont {R.}~\bibnamefont {Gendriesch}}, \bibinfo
  {author} {\bibfnamefont {S.~P.}\ \bibnamefont {Belov}}, \ and\ \bibinfo
  {author} {\bibfnamefont {G.}~\bibnamefont {Winnewisser}},\ }\href@noop {}
  {\bibfield  {journal} {\bibinfo  {journal} {J. Mol. Spectrosc.}\ }\textbf
  {\bibinfo {volume} {201}},\ \bibinfo {pages} {124} (\bibinfo {year}
  {2000})}\BibitemShut {NoStop}%
\bibitem [{\citenamefont {Jongma}\ \emph {et~al.}(1995)\citenamefont {Jongma},
  \citenamefont {Rasing},\ and\ \citenamefont {Meijer}}]{Jongma:JCP}%
  \BibitemOpen
  \bibfield  {author} {\bibinfo {author} {\bibfnamefont {R.~T.}\ \bibnamefont
  {Jongma}}, \bibinfo {author} {\bibfnamefont {T.}~\bibnamefont {Rasing}}, \
  and\ \bibinfo {author} {\bibfnamefont {G.}~\bibnamefont {Meijer}},\
  }\href@noop {} {\bibfield  {journal} {\bibinfo  {journal} {J. Chem. Phys.}\
  }\textbf {\bibinfo {volume} {102}},\ \bibinfo {pages} {1925} (\bibinfo {year}
  {1995})}\BibitemShut {NoStop}%
\bibitem [{\citenamefont {Hannemann}\ \emph
  {et~al.}(2006{\natexlab{a}})\citenamefont {Hannemann}, \citenamefont
  {Salumbides}, \citenamefont {Witte}, \citenamefont {Zinkstok}, \citenamefont
  {van Duijn}, \citenamefont {Eikema},\ and\ \citenamefont
  {Ubachs}}]{Hannemann:PRA1}%
  \BibitemOpen
  \bibfield  {author} {\bibinfo {author} {\bibfnamefont {S.}~\bibnamefont
  {Hannemann}}, \bibinfo {author} {\bibfnamefont {E.~J.}\ \bibnamefont
  {Salumbides}}, \bibinfo {author} {\bibfnamefont {S.}~\bibnamefont {Witte}},
  \bibinfo {author} {\bibfnamefont {R.~T.}\ \bibnamefont {Zinkstok}}, \bibinfo
  {author} {\bibfnamefont {E.~J.}\ \bibnamefont {van Duijn}}, \bibinfo {author}
  {\bibfnamefont {K.~S.~E.}\ \bibnamefont {Eikema}}, \ and\ \bibinfo {author}
  {\bibfnamefont {W.}~\bibnamefont {Ubachs}},\ }\href@noop {} {\bibfield
  {journal} {\bibinfo  {journal} {Phys. Rev. A}\ }\textbf {\bibinfo {volume}
  {74}},\ \bibinfo {pages} {062514} (\bibinfo {year}
  {2006}{\natexlab{a}})}\BibitemShut {NoStop}%
\bibitem [{\citenamefont {Hannemann}\ \emph
  {et~al.}(2006{\natexlab{b}})\citenamefont {Hannemann}, \citenamefont
  {Salumbides}, \citenamefont {Witte}, \citenamefont {Zinkstok}, \citenamefont
  {van Duijn}, \citenamefont {Eikema},\ and\ \citenamefont
  {Ubachs}}]{Hannemann:PRA2}%
  \BibitemOpen
  \bibfield  {author} {\bibinfo {author} {\bibfnamefont {S.}~\bibnamefont
  {Hannemann}}, \bibinfo {author} {\bibfnamefont {E.~J.}\ \bibnamefont
  {Salumbides}}, \bibinfo {author} {\bibfnamefont {S.}~\bibnamefont {Witte}},
  \bibinfo {author} {\bibfnamefont {R.~T.}\ \bibnamefont {Zinkstok}}, \bibinfo
  {author} {\bibfnamefont {E.~J.}\ \bibnamefont {van Duijn}}, \bibinfo {author}
  {\bibfnamefont {K.~S.~E.}\ \bibnamefont {Eikema}}, \ and\ \bibinfo {author}
  {\bibfnamefont {W.}~\bibnamefont {Ubachs}},\ }\href@noop {} {\bibfield
  {journal} {\bibinfo  {journal} {Phys. Rev. A}\ }\textbf {\bibinfo {volume}
  {74}},\ \bibinfo {pages} {012505} (\bibinfo {year}
  {2006}{\natexlab{b}})}\BibitemShut {NoStop}%
\bibitem [{\citenamefont {Hannemann}\ \emph {et~al.}(2007)\citenamefont
  {Hannemann}, \citenamefont {Salumbides},\ and\ \citenamefont
  {Ubachs}}]{Hannemann:OptLett}%
  \BibitemOpen
  \bibfield  {author} {\bibinfo {author} {\bibfnamefont {S.}~\bibnamefont
  {Hannemann}}, \bibinfo {author} {\bibfnamefont {E.~J.}\ \bibnamefont
  {Salumbides}}, \ and\ \bibinfo {author} {\bibfnamefont {W.}~\bibnamefont
  {Ubachs}},\ }\href@noop {} {\bibfield  {journal} {\bibinfo  {journal} {Opt.
  Lett.}\ }\textbf {\bibinfo {volume} {32}},\ \bibinfo {pages} {1381} (\bibinfo
  {year} {2007})}\BibitemShut {NoStop}%
\bibitem [{\citenamefont {Groenenboom}()}]{groenenboomprivatecommunications}%
  \BibitemOpen
  \bibfield  {author} {\bibinfo {author} {\bibfnamefont {G.~C.}\ \bibnamefont
  {Groenenboom}},\ }\href@noop {} {}\bibinfo {howpublished} {private
  communication}\BibitemShut {NoStop}%
\bibitem [{\citenamefont {Minaev}\ \emph {et~al.}(1995)\citenamefont {Minaev},
  \citenamefont {Plachkevytch},\ and\ \citenamefont {{\AA}gren}}]{Minaev}%
  \BibitemOpen
  \bibfield  {author} {\bibinfo {author} {\bibfnamefont {B.}~\bibnamefont
  {Minaev}}, \bibinfo {author} {\bibfnamefont {O.}~\bibnamefont
  {Plachkevytch}}, \ and\ \bibinfo {author} {\bibfnamefont {H.}~\bibnamefont
  {{\AA}gren}},\ }\href@noop {} {\bibfield  {journal} {\bibinfo  {journal} {J.
  Chem. Soc. Faraday Trans.}\ }\textbf {\bibinfo {volume} {91}},\ \bibinfo
  {pages} {1729} (\bibinfo {year} {1995})}\BibitemShut {NoStop}%
\bibitem [{\citenamefont {Jongma}\ \emph {et~al.}(1997)\citenamefont {Jongma},
  \citenamefont {von Helden}, \citenamefont {Berden},\ and\ \citenamefont
  {Meijer}}]{Jongma:CPL}%
  \BibitemOpen
  \bibfield  {author} {\bibinfo {author} {\bibfnamefont {R.~T.}\ \bibnamefont
  {Jongma}}, \bibinfo {author} {\bibfnamefont {G.}~\bibnamefont {von Helden}},
  \bibinfo {author} {\bibfnamefont {G.}~\bibnamefont {Berden}}, \ and\ \bibinfo
  {author} {\bibfnamefont {G.}~\bibnamefont {Meijer}},\ }\href@noop {}
  {\bibfield  {journal} {\bibinfo  {journal} {Chem. Phys. Lett.}\ }\textbf
  {\bibinfo {volume} {270}},\ \bibinfo {pages} {304} (\bibinfo {year}
  {1997})}\BibitemShut {NoStop}%
\bibitem [{\citenamefont {Witte}\ \emph {et~al.}(2005)\citenamefont {Witte},
  \citenamefont {Zinkstok}, \citenamefont {Ubachs}, \citenamefont
  {Hogervorst},\ and\ \citenamefont {Eikema}}]{Witte:Science}%
  \BibitemOpen
  \bibfield  {author} {\bibinfo {author} {\bibfnamefont {S.}~\bibnamefont
  {Witte}}, \bibinfo {author} {\bibfnamefont {R.~T.}\ \bibnamefont {Zinkstok}},
  \bibinfo {author} {\bibfnamefont {W.}~\bibnamefont {Ubachs}}, \bibinfo
  {author} {\bibfnamefont {W.}~\bibnamefont {Hogervorst}}, \ and\ \bibinfo
  {author} {\bibfnamefont {K.~S.~E.}\ \bibnamefont {Eikema}},\ }\href@noop {}
  {\bibfield  {journal} {\bibinfo  {journal} {Science}\ }\textbf {\bibinfo
  {volume} {307}},\ \bibinfo {pages} {400} (\bibinfo {year}
  {2005})}\BibitemShut {NoStop}%
\bibitem [{\citenamefont {Salumbides}\ \emph {et~al.}(2008)\citenamefont
  {Salumbides}, \citenamefont {Bailly}, \citenamefont {Khramov}, \citenamefont
  {Wolf}, \citenamefont {Eikema}, \citenamefont {Vervloet},\ and\ \citenamefont
  {Ubachs}}]{Salumbides}%
  \BibitemOpen
  \bibfield  {author} {\bibinfo {author} {\bibfnamefont {E.~J.}\ \bibnamefont
  {Salumbides}}, \bibinfo {author} {\bibfnamefont {D.}~\bibnamefont {Bailly}},
  \bibinfo {author} {\bibfnamefont {A.}~\bibnamefont {Khramov}}, \bibinfo
  {author} {\bibfnamefont {A.~L.}\ \bibnamefont {Wolf}}, \bibinfo {author}
  {\bibfnamefont {K.~S.~E.}\ \bibnamefont {Eikema}}, \bibinfo {author}
  {\bibfnamefont {M.}~\bibnamefont {Vervloet}}, \ and\ \bibinfo {author}
  {\bibfnamefont {W.}~\bibnamefont {Ubachs}},\ }\href@noop {} {\bibfield
  {journal} {\bibinfo  {journal} {Phys. Rev. Lett.}\ }\textbf {\bibinfo
  {volume} {101}},\ \bibinfo {pages} {223001} (\bibinfo {year}
  {2008})}\BibitemShut {NoStop}%
\bibitem [{\citenamefont {Brown}\ and\ \citenamefont
  {Merer}(1979)}]{BrownMerer}%
  \BibitemOpen
  \bibfield  {author} {\bibinfo {author} {\bibfnamefont {J.~M.}\ \bibnamefont
  {Brown}}\ and\ \bibinfo {author} {\bibfnamefont {A.~J.}\ \bibnamefont
  {Merer}},\ }\href@noop {} {\bibfield  {journal} {\bibinfo  {journal} {J. Mol.
  Spectrosc.}\ }\textbf {\bibinfo {volume} {74}},\ \bibinfo {pages} {488}
  (\bibinfo {year} {1979})}\BibitemShut {NoStop}%
\bibitem [{\citenamefont {Western}(2009)}]{pgopher}%
  \BibitemOpen
  \bibfield  {author} {\bibinfo {author} {\bibfnamefont {C.~M.}\ \bibnamefont
  {Western}},\ }\href@noop {} {\emph {\bibinfo {title} {PGOPHER, a Program for
  Simulating Rotational Structure, Version 6.0.111}}},\ \bibinfo {organization}
  {University of Bristol} (\bibinfo {year} {2009}),\ \bibinfo {note}
  {http://pgopher.chm.bris.ac.uk}\BibitemShut {NoStop}%
\bibitem [{\citenamefont {Winnewisser}\ \emph {et~al.}(1997)\citenamefont
  {Winnewisser}, \citenamefont {Belov}, \citenamefont {Klaus},\ and\
  \citenamefont {Schieder}}]{Winnewisser:1997}%
  \BibitemOpen
  \bibfield  {author} {\bibinfo {author} {\bibfnamefont {G.}~\bibnamefont
  {Winnewisser}}, \bibinfo {author} {\bibfnamefont {S.~P.}\ \bibnamefont
  {Belov}}, \bibinfo {author} {\bibfnamefont {T.}~\bibnamefont {Klaus}}, \ and\
  \bibinfo {author} {\bibfnamefont {R.}~\bibnamefont {Schieder}},\ }\href@noop
  {} {\bibfield  {journal} {\bibinfo  {journal} {J. Mol. Spectrosc.}\ }\textbf
  {\bibinfo {volume} {184}},\ \bibinfo {pages} {468} (\bibinfo {year}
  {1997})}\BibitemShut {NoStop}%
\bibitem [{\citenamefont {Brown}\ \emph {et~al.}(1977)\citenamefont {Brown},
  \citenamefont {Kopp}, \citenamefont {Malmberg},\ and\ \citenamefont
  {Rydh}}]{Brown:1977}%
  \BibitemOpen
  \bibfield  {author} {\bibinfo {author} {\bibfnamefont {J.~M.}\ \bibnamefont
  {Brown}}, \bibinfo {author} {\bibfnamefont {I.}~\bibnamefont {Kopp}},
  \bibinfo {author} {\bibfnamefont {C.}~\bibnamefont {Malmberg}}, \ and\
  \bibinfo {author} {\bibfnamefont {B.}~\bibnamefont {Rydh}},\ }\href@noop {}
  {\bibfield  {journal} {\bibinfo  {journal} {Phys. Scripta}\ }\textbf
  {\bibinfo {volume} {17}},\ \bibinfo {pages} {55} (\bibinfo {year}
  {1977})}\BibitemShut {NoStop}%
\bibitem [{\citenamefont {Puzzarini}\ \emph {et~al.}(2003)\citenamefont
  {Puzzarini}, \citenamefont {Dore},\ and\ \citenamefont
  {Cazzoli}}]{Puzzarini:2003}%
  \BibitemOpen
  \bibfield  {author} {\bibinfo {author} {\bibfnamefont {C.}~\bibnamefont
  {Puzzarini}}, \bibinfo {author} {\bibfnamefont {L.}~\bibnamefont {Dore}}, \
  and\ \bibinfo {author} {\bibfnamefont {G.}~\bibnamefont {Cazzoli}},\
  }\href@noop {} {\bibfield  {journal} {\bibinfo  {journal} {J. Mol.
  Spectrosc.}\ }\textbf {\bibinfo {volume} {217}},\ \bibinfo {pages} {19}
  (\bibinfo {year} {2003})}\BibitemShut {NoStop}%
\bibitem [{\citenamefont {Coxon}\ and\ \citenamefont
  {Hajigeorgiou}(2004)}]{Coxon}%
  \BibitemOpen
  \bibfield  {author} {\bibinfo {author} {\bibfnamefont {J.~A.}\ \bibnamefont
  {Coxon}}\ and\ \bibinfo {author} {\bibfnamefont {P.~G.}\ \bibnamefont
  {Hajigeorgiou}},\ }\href@noop {} {\bibfield  {journal} {\bibinfo  {journal}
  {J. Chem. Phys.}\ }\textbf {\bibinfo {volume} {121}},\ \bibinfo {pages}
  {2992} (\bibinfo {year} {2004})}\BibitemShut {NoStop}%
\bibitem [{\citenamefont {Ubachs}\ \emph {et~al.}(2000)\citenamefont {Ubachs},
  \citenamefont {Velchev},\ and\ \citenamefont {Cacciani}}]{Ubachs:2000}%
  \BibitemOpen
  \bibfield  {author} {\bibinfo {author} {\bibfnamefont {W.}~\bibnamefont
  {Ubachs}}, \bibinfo {author} {\bibfnamefont {I.}~\bibnamefont {Velchev}}, \
  and\ \bibinfo {author} {\bibfnamefont {P.}~\bibnamefont {Cacciani}},\
  }\href@noop {} {\bibfield  {journal} {\bibinfo  {journal} {J. Chem. Phys.}\
  }\textbf {\bibinfo {volume} {113}},\ \bibinfo {pages} {547} (\bibinfo {year}
  {2000})}\BibitemShut {NoStop}%
\bibitem [{\citenamefont {Cacciani}\ \emph {et~al.}(2001)\citenamefont
  {Cacciani}, \citenamefont {Brandi}, \citenamefont {Velchev}, \citenamefont
  {Lyng{\aa}}, \citenamefont {Wahlstr\"om},\ and\ \citenamefont
  {Ubachs}}]{Ubachs:2001}%
  \BibitemOpen
  \bibfield  {author} {\bibinfo {author} {\bibfnamefont {P.}~\bibnamefont
  {Cacciani}}, \bibinfo {author} {\bibfnamefont {F.}~\bibnamefont {Brandi}},
  \bibinfo {author} {\bibfnamefont {I.}~\bibnamefont {Velchev}}, \bibinfo
  {author} {\bibfnamefont {C.}~\bibnamefont {Lyng{\aa}}}, \bibinfo {author}
  {\bibfnamefont {C.~G.}\ \bibnamefont {Wahlstr\"om}}, \ and\ \bibinfo {author}
  {\bibfnamefont {W.}~\bibnamefont {Ubachs}},\ }\href@noop {} {\bibfield
  {journal} {\bibinfo  {journal} {Eur. Phys. J.D.}\ }\textbf {\bibinfo {volume}
  {15}},\ \bibinfo {pages} {47} (\bibinfo {year} {2001})}\BibitemShut {NoStop}%
\bibitem [{\citenamefont {{Le Roy}}(2007)}]{level}%
  \BibitemOpen
  \bibfield  {author} {\bibinfo {author} {\bibfnamefont {R.~J.}\ \bibnamefont
  {{Le Roy}}},\ }\href@noop {} {\emph {\bibinfo {title} {LEVEL 8.0: A Computer
  Program for Solving the Radial Schr\"odinger Equation for Bound and
  Quasibound Levels}}},\ \bibinfo {organization} {University of Waterloo
  Chemical Physics Research Report CP-663} (\bibinfo {year} {2007}),\ \bibinfo
  {note} {see http://leroy.uwaterloo.ca/programs/}\BibitemShut {NoStop}%
\bibitem [{\citenamefont {{Le Roy}}()}]{LeRoyprivatecommunications}%
  \BibitemOpen
  \bibfield  {author} {\bibinfo {author} {\bibfnamefont {R.~J.}\ \bibnamefont
  {{Le Roy}}},\ }\href@noop {} {}\bibinfo {howpublished} {private
  communication}\BibitemShut {NoStop}%
\bibitem [{\citenamefont {Angeli}(1998)}]{Angeli}%
  \BibitemOpen
  \bibfield  {author} {\bibinfo {author} {\bibfnamefont {I.}~\bibnamefont
  {Angeli}},\ }\href@noop {} {\bibfield  {journal} {\bibinfo  {journal} {Acta
  Phys. Hung. New Ser.-Heavy Ion Phys.}\ }\textbf {\bibinfo {volume} {8}},\
  \bibinfo {pages} {23} (\bibinfo {year} {1998})}\BibitemShut {NoStop}%
\bibitem [{\citenamefont {Flambaum}\ \emph {et~al.}(2004)\citenamefont
  {Flambaum}, \citenamefont {Leinweber}, \citenamefont {Thomas},\ and\
  \citenamefont {Young}}]{Flambaum:2004}%
  \BibitemOpen
  \bibfield  {author} {\bibinfo {author} {\bibfnamefont {V.~V.}\ \bibnamefont
  {Flambaum}}, \bibinfo {author} {\bibfnamefont {D.~B.}\ \bibnamefont
  {Leinweber}}, \bibinfo {author} {\bibfnamefont {A.~W.}\ \bibnamefont
  {Thomas}}, \ and\ \bibinfo {author} {\bibfnamefont {R.~D.}\ \bibnamefont
  {Young}},\ }\href@noop {} {\bibfield  {journal} {\bibinfo  {journal} {Phys.
  Rev. D}\ }\textbf {\bibinfo {volume} {69}},\ \bibinfo {pages} {115006}
  (\bibinfo {year} {2004})}\BibitemShut {NoStop}%
\bibitem [{\citenamefont {Dent}(2007)}]{Dent:2007}%
  \BibitemOpen
  \bibfield  {author} {\bibinfo {author} {\bibfnamefont {T.}~\bibnamefont
  {Dent}},\ }\href@noop {} {\bibfield  {journal} {\bibinfo  {journal} {J.
  Cosmol. Astropart. Phys.}\ }\textbf {\bibinfo {volume} {1}},\ \bibinfo
  {pages} {13} (\bibinfo {year} {2007})}\BibitemShut {NoStop}%
\bibitem [{Note1()}]{Note1}%
  \BibitemOpen
  \bibinfo {note} {Note that Bethlem and Ubachs~\cite {Bethlem:2009}
  erroneously used $K_\mu =A/\nu $ instead of $K_\mu =A/2\nu $.}\BibitemShut
  {Stop}%
\end{thebibliography}

\providecommand{\noopsort}[1]{}\providecommand{\singleletter}[1]{#1}%

\end{document}